\documentclass[aps,floatfix,preprint,nofootinbib]{revtex4}
\usepackage{cancel}
\usepackage{graphicx}
\usepackage{amssymb}
\usepackage{epstopdf}
\usepackage{verbatim}
\newcommand{\mbeq}{\stackrel{!}{=}}
\begin{document}

\title{Implications of the Higgs Boson and the LHC for the MSSM}

\author{A. Pierce}

\affiliation{Michigan Center for Theoretical Physics, \\
Department of Physics, \\
University of Michigan,\\
Ann Arbor, MI 48109, USA}
\email{atpierce@umich.edu}



\begin{abstract}
These lectures were presented at the TASI 2012 summer school to a mixture of graduate students in particle theory and cosmology.  They serve as an elementary introduction to the Minimal Supersymmetric Standard Model (MSSM) and discuss the implications of a 125 GeV Higgs boson for this theory.  Some familiarity with the Standard Model is assumed.  
\end{abstract}

\maketitle

\newpage

\section{Introduction}\label{AP:sec:intro}

As recently as 2008, direct experimental knowledge of the Higgs boson was limited.  All that was known was that its mass lay above the LEP bound \cite{LEPBound} of 114 GeV. But progress has been rapid.  At the beginning of this decade, the Tevatron and the Large Hadron Collider (LHC) quickly began to rule out large swathes of potential Higgs boson masses, and at the time these lectures were delivered, there were tantalizing hints of a Higgs boson at 125 GeV\cite{DecHintATLAS,DecHintCMS}.   Just a couple of weeks later, these hints were solidified by a discovery of a ``Higgs-like" state near 125 GeV by both the ATLAS and CMS experiments\cite{JulyDiscCMS,JulyDiscATLAS}.  Around the same time, the Tevatron experiments also presented evidence for a Higgs-like state decaying to $b$ quarks\cite{Tevatronbb}.  Working under the assumption that this state is in fact a Higgs boson, what are the implications for supersymmetry?

These lectures begin with an introduction to supersymmetry in two parts. First, we present a discussion of why we are interested in weak scale supersymmetry in the first place.  With our appetites whetted, we move to overview the basic structure of the Minimal Supersymmetric Standard Model (MSSM).  The treatment is a quick and dirty one.  The goal is  to give non-experts some appreciation for the structure of the MSSM, how to look for it, and how a 125 GeV Higgs boson fits into the picture.  Because of our interest in the Higgs signal, we analyze the MSSM Higgs potential in some detail, giving much of the rest of the model short shrift.  The motivated student will want to consult some of the several excellent reviews of this subject for more details.  An incomplete list of recent pedagogical resources includes lecture notes, Refs.~\cite{MartinPrimer, Martin:2012us,DreesIntro, DineTASI, LykkenTASI, BaggerTASI, HitoshiNotes}, as well as textbooks, Refs~\cite{WessBagger,DreesText, WeinbergText, Binetruy:2006ad, DineBook, Terning:2006bq}.  

\section{Why supersymmetry?}
\label{sec:WhySUSY}
Before targeting supersymmetry, it is illuminating to first ask: why go beyond the Standard Model (SM) at all?  

There are experimental reasons to extend the Standard Model.  Some new particle must play the role of the Dark Matter which makes up the majority of mass in our universe.  In addition, the Standard Model alone cannot explain the asymmetry between matter and antimatter in our universe.  Moreover, if an epoch of inflation occurred early in the history of the universe, as is suggested by data from the Cosmic Microwave Background (CMB), there must be new physics to drive that inflation.  All these observations warrant new physics.   Unfortunately, they do not indicate the energy scale at which the new physics arises.  It could be near the weak scale $M_W$ (in which case we are on the cusp of probing it), or be stubbornly out of reach. 

There are also purely aesthetic reasons to go beyond the Standard Model.  
For example, the Standard Model may have too many free parameters for comfort, some of which display strange patterns. Perhaps some semblance of order might be restored if the seemingly disparate forces could be satisfyingly unified into a single force, but presumably again at much higher energies, far removed from $M_{W}$.
 And there is the puzzle  puzzle of why such disparate values of fermion masses are present within the Standard Model (e.g., the electron mass is six orders of magnitude smaller than that of the top quark).  But again, there is no guarantee that the explanation -- and the attendant new physics -- should lie near the weak scale.  
  
 Of all the motivations to go beyond the Standard Model, both experimental and aesthetic, there is only one that firmly points to new physics at the weak scale: why is gravity so much weaker than the other forces? Equivalently, how can we explain the hierarchy  $M_W << M_{pl}$?   As we will discuss below, if the Standard Model model remains unmodified to high scales, the chasm between the weak scale, $M_{W} \sim 100 $ GeV and the scale associated with gravity, $M_{pl} \sim 10^{18}$ GeV is unexpected.    Quantum corrections attempt to drag the Higgs boson mass up to the highest energy scale at which the Standard Model remains valid.  If this is the Planck scale, then we must ask why the Higgs boson stubbornly remains light in the face of these extreme quantum pressures.   This is known as the fine--tuning or the hierarchy problem. The simplest solution is that the Standard Model is not, in fact, valid up to very high energies, and new physics arrives close to the weak scale to tame these quantum corrections.

But not just any new physics can play Siegfried and Roy to these quantum lions.  But as we discuss below, supersymmetry is an example of the kind of physics that fits the bill, and this is one of the main reasons it is touted as an extension to the SM.  In the MSSM the couplings of new particles conspire (because of the symmetry) so that the quantum pressures from these new particles precisely cancel those that would have dragged the Higgs mass Planckward.  But not without a price: supersymmetry posits the doubling of the Standard Model particles, an no such partners  have been observed as yet.   

\subsection{Supersymmetry isn't crazy}
This price may seem awfully high, and might  cause one to dismiss supersymmetry out of hand.  Here we present a historical analogy (which we believe first appeared in Ref.~\cite{HitoshiSUSY}) that helps illustrate why such a doubling might not be so outlandish.  Suppose, armed with knowledge of classical electromagnetism, one set about trying to calculate the correction to the electron's rest mass due to its electrical self energy.  The result is parametrically
\begin{eqnarray}
\label{eqn:ClassicalElectron}
\Delta E_{elec} &\sim& -\frac{\alpha}{r_{e}}, \nonumber \\
&=& -1.4 \, \rm{MeV} \left( \frac{\rm{fm}}{r_{e}} \right)
\end{eqnarray}
where the $r_{e}$ represents a ``radius of the electron".  It is not {\it a priori} clear what value one should choose for this electron radius.  With detailed knowledge about the lack of electron substructure from LEP, one might choose something like something like $(\rm{TeV})^{-1}$ or even smaller.  (This is a bit artificial, because these experimental limits depend upon on a knowledge of Quantum Field Theory, and we are imagining a strictly classical calculation, but let us see where this takes us.)  Plugging in this value to Eq.~(\ref{eqn:ClassicalElectron}),  one would find a contribution to the \emph{electron mass}  (every electron has it after all) of -10 GeV.   Comparing to its physical value, one would be forced to conclude that nature has engaged in a delicate balancing act:
\begin{equation}
\label{eq:ElectronFineTune}
m_e c^{2} = .511 \, {\rm MeV} =  -10 \; {\rm GeV} + 10.000511 \; {\rm GeV} = \Delta E_{elec} + m _{0} c^{2}
\end{equation}
where $m_{0} c^{2}$ represents all other contributions to the electron mass (including, e.g., the bare value).   

But this calculation is incorrect -- there is a sense in which the classical electron is in fact ``large" on the TeV scale.  Well before we get to these energies, the existence of the positron becomes important.  As first calculated by Victor Weisskopf, once its contributions to the self-energy are taken into account, we find:
\begin{equation}
m= m_{e}^{bare} \left( 1 + \frac{3 \alpha}{4 \pi} \log{\left(\frac{\frac{\hbar}{m_{e}c}}{R} \right)} + \cdots \right)
\end{equation}
Here $\frac{\hbar}{m_{e}c} \approx 4 \times 10^{-13}$ cm, and $R$ is a new length scale that tells us where this improved calculation breaks down.  Even for $R \sim 1/M_{pl}$ this is only a 10\% correction.
 
Thus the existence of partner particles (here the positron postulated by Dirac in 1928, and found by Andersen in 1932) was crucial to the cancellation.  Furthermore, these particles are intimately tied to a symmetry (here the Poincar\'e symmetry that allows the formulation of a chiral symmetry).  So, a symmetry forced nature to  ``double" the number of particles, and  the new partner particles softened the self-energy.   Is it possible that this story repeats itself in nature, with supersymmetry playing the role of the doubling symmetry?

\subsection{How would this apply to the Standard Model?}
The fermions of the Standard Model (thanks to the chiral symmetry discussed above) do not require a delicate balancing act in their mass budget.  However, the situation is different for scalars, and in particular, the Standard Model Higgs boson, whose vacuum expectation value sets the electroweak scale.  Consider the Standard Model Higgs potential: 
\begin{equation}
V(H) = -\mu^{2} |H|^{2} + \lambda |H|^{4}
\end{equation}
with 
\begin{equation}
\langle H \rangle = \frac{1}{\sqrt{2}} \left(\begin{array}{c} 0 \\ v \end{array} \right)
\hspace{.5in} v = \sqrt{\frac{\mu^{2}}{\lambda}} \hspace{.5in} v = 246 \, \rm{GeV}.
\end{equation}
Just as the classical electron received large corrections to its mass electrical interactions, the Higgs field receives large corrections to its mass squared parameter $\mu^{2}$ via, e.g., its coupling to the top quark.  

\begin{figure}
\begin{center}
\includegraphics[width=0.6\textwidth]{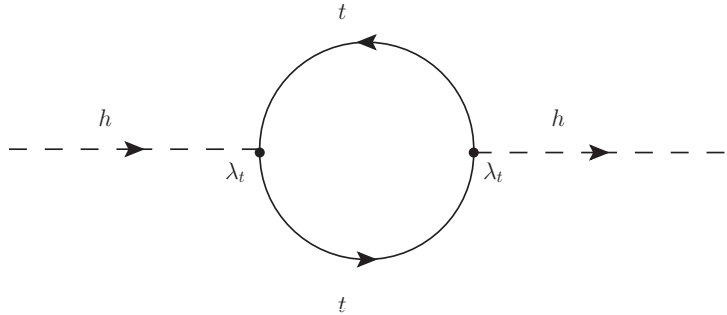}
\end{center}
\caption{Leading correction to the Standard Model Higgs mass squared parameter from the top quark.}
\label{fig:TopLoop} 
\end{figure}

The diagram shown in Fig.~\ref{fig:TopLoop} 
contributes a value
\begin{equation}
\Delta (\mu^{2})^{1-loop}_{top}  = -\frac{3 \lambda_{t}^2}{8 \pi^2} \Lambda^2 ,
\end{equation}
where $\lambda_{t}$ is the Yukawa coupling to the top quark, and $\Lambda$ corresponds to the scale where our calculation breaks down.  If, for example, $\Lambda$ were identified with $M_{pl}$ -- as would be appropriate if the Standard Model were valid all the way up to scales were gravity becomes strong --  then  we have an equation of the form:
\begin{equation}
\label{eqn:SMft}
\mu_{0}^2 + \Delta \mu^{2} \sim M_{W}^2
\end{equation}
where $\Delta \mu^{2}$ is about $10^{32}$ times $M_{W}^2$!  This makes the conspiracy of Eq.~(\ref{eq:ElectronFineTune}) seem tame.  Such a fine-tuning is said to be ``unnatural".  The desire for naturalness in the physics of the electroweak scale is the main motivator for new physics at the LHC, for the way to avoid an unnatural theory is to posit that $\Lambda$ is not $M_{pl}$ but instead a much lower scale, of order a TeV.  Then $\Delta \mu^{2}$ will be the same size as $M_{W}$.    

Perhaps we can gain insight on fine-tuning by considering an example outside the realm of particle physics:  the Landau-Ginzburg theory for superconductors.  There the free energy takes a form
\begin{equation}
F = \alpha |\phi|^2 + \beta |\phi|^4,
\end{equation}
with $\alpha$ having leading temperature dependence
\begin{equation}
\alpha(T) = \alpha_{0} (T-T_C),
\end{equation}
where $T_{c}$ is the critical temperature corresponding to the phase transition.   For $T < T_{c}$, minimization of the free energy leads to a non-zero vacuum expectation value of $\phi$.  This non-zero value is the ``order parameter" associated with the transition to the superconducting state.  Furthermore, below the transition point, the size of the order parameter is proportional to the distance from the critical temperature.   One would expect $\alpha_0$ to be related to the fundamental mass scales in the theory.  If the Standard Model is unchanged up to the Planck scale, the incredible smallness of the electroweak scale relative to the Planck scale is like having a superconducting system tuned to a temperature extraordinary close to the transition temperature.  Moreover, the Higgs vacuum expectation is pegged there in spite of the quantum fluctuations (cf. thermal fluctuations in the superconductor case).   In the case of the superconductor, it is clear what is doing the fine-tuning -- it is the experimentalist, who is diligently keeping her sample within a hair's breadth of the critical temperature.  But what physics is doing that job for the electroweak theory?


\subsection{A bit more on Naturalness}
This discussion about naturalness and fine-tuning may seem abstract.  Given its importance for determining whether we will, in fact, see new physics at the LHC, it would be useful to know whether or not there are examples where naturalness is a useful guiding principle.  A lucid discussion of these points appears in Ref.~\cite{Giudice:2008bi}.  

One example is the mass splitting between the charged and neutral pion.  This mass difference receives a contribution from the the the electromagnetic interaction, and is logarithmically divergent.   This divergence is effectively cut-off by the $\rho$-meson mass, so that the size of the contribution to the mass splitting is ``natural".    The mass of the $\rho$ was not predicted in this way, but it is indeed a natural system.  

Another, perhaps the most compelling, historical example \cite{GaillardLee} relates to the discovery of the charm quark.  If one calculates in an effective theory that has only $u,d,s$ quarks and weak interactions, the contribution to the neutral kaon mass splitting is quadratically sensitive to the cutoff of the effective theory $\Lambda$:
\begin{equation}
\frac{\Delta M_{K}}{M_{K^0_L}} = \frac{G_{F} f_{K}}{6 \pi^{2}} \sin^2 \theta_{C} \Lambda^{2},
\end{equation}
with $\Delta M_{K} \equiv M_{K^0_L}-M_{K^0_S}$, $G_{F}$ the Fermi constant, and $f_{K} = 114$ MeV the Kaon decay constant.   The experimentally observed value \cite{PDG} of the left hand side is $7 \times 10^{-15}$.  Requiring a natural theory imposes that the right-hand-side not supersaturate this value.  This is the condition that the contribution calculated in the effective theory need not be finely tuned against other contributions.  Here this  implies that the cutoff $\Lambda < 2$ GeV.  So, naturalness indicates that new physics should arise before this scale, and in fact it does:  the charm quark  has mass $m_{c} =1.2$ GeV.  Its existence is the foundation of the  Glashow-Illiopoulos-Maini (GIM) mechanism.  

Whether the weak scale is in fact natural is an experimental question -- one that we should the answer to soon.

\section{A shortcut to building the MSSM}
We have foreshadowed that the doubling of particles in the MSSM will soften the ultraviolet behavior and reduce the fine-tuning.  This results from a cancellation between loops of fermions (like the top quark) and bosons (like its superpartner, the stop).   It is the supersymmetry itself that assures this cancelation.   To appreciate why this is so, we first need to understand something about the interactions in these theories. 

We take a relatively streamlined approach to quickly get to the structure of the MSSM, glossing over important details and stating some results without proof. For details, consult Ref.~\cite{MartinPrimer}, or other lecture notes referred to in the introduction.

\subsection{Some Basics of Supersymmetric Theories}
\label{sec:SUSYBasics}

In supersymmetric theories, particles can be packaged into ``supermultiplets" that serve as building blocks for the theory.  Particles within a supermultiplet share gauge charges but differ in spin.  Using these building blocks, we can find a compact presentation of the MSSM.  Since supersymmetry relates fermions and bosons and supermultiplets contain both, they allow us to keep supersymmetry manifest throughout the construction.  For the MSSM we will focus on two multiplet types:\footnote{
Here,  for the most part we will ignore gravity, which requires a multiplet of its own.  If the gravitino, the superpartner of the graviton is the lightest superpartner is the LSP, there can be important impacts on collider phenomenology.  For a brief overview, see Ref.~\cite{MartinPrimer}.}
\begin{itemize}
\item{Chiral multiplets, which contain a complex scalar and a Weyl fermion.}
\item{Vector multiplets, which contain  a vector field and a fermion (gaugino).}
\end{itemize}
We will first introduce these multiplets, and then describe how to use them to construct the supersymmetric Lagrangian of the MSSM.

Precisely because supersymmetry relates bosons and fermions, in each multiplet the bosonic and fermionic degrees of freedom must match.  In the chiral multiplet, two real bosonic degrees of freedom are bundled as a complex scalar, $\phi$. 
The chiral superfield $\Phi$ can  be written as
\begin{equation}
\label{eq:chiralSF}
\Phi(y) = \phi(y) + \sqrt{2} \theta \psi(y) + \theta^{2} F(y),
\end{equation}
with $\psi$ a Weyl Fermion.  $F$ is an auxiliary field (whose purpose we discuss below), and
\begin{equation}
\label{eq:ydef}
y^{\mu} \equiv x^{\mu} - i \theta \sigma^{\mu} \bar{\theta}.
\end{equation}
 Each SM matter field (i.e. quark or lepton) gets promoted to a chiral multiplet and will be identified with these $\psi$ --- meaning that the MSSM will have scalar partners for each field.  The names of scalar superpartners are found by  prepending an ``s" (for scalar) to the Standard Model particle name, e.g. ``squark" and ``slepton".  (The names of fermionic superpartners are found by appending the suffix ``ino", e.g. ``Higgsino" or ``gaugino".) The $\theta$ that appears here is a two component Grassman (anti-commuting) spinor.  The product of $\psi$ and $\theta$ is bosonic, as it should be -- everything on the RHS of Eq.~(\ref{eq:chiralSF}) should have the same statistics.  Because $\theta$ is anti-commuting and only has two independent components,  $\theta^{n}  = 0$, for $n \geq 3$.  A consequence is that Taylor expansions (about $\theta =0$) quickly truncate.  A Weyl fermion has two degrees of freedom, so this balances against the two bosonic degrees of freedom  in $\phi$.  From studies of QED,  four component fermions (and Dirac matrices flying fast and furiously) may be more familiar, but the Standard Model electroweak gauge group treats left-handed and right-handed fermions differently:  it is a chiral gauge theory.  When dealing with such theories -- in contrast to the vectorlike theories of QCD and QED -- it is more natural to use two component notation (and it is Pauli spin matrices do the flying).\footnote{The conversion between these two languages can be a little tricky at first if you are not used to it.  The recent treatise by Dreiner, Haber and Martin \cite{Dreiner:2008tw} is a comprehensive and useful reference on two-component notation, and includes a nice appendix on how to go back and forth to four-component notation.}  

We now return to discuss the auxiliary field $F$.  Off-shell a Weyl fermion has four, rather than two degrees of freedom.  How, then, does the accounting work in the superfield to preserve the symmetry between bosons and fermions?  The answer is to introduce a bosonic auxiliary field with two components of its own.  It is non-propagating, and on-shell it will just be a function of the other scalar fields in the theory.   Off-shell, it provides two additional bosonic degrees of freedom, and supersymmetry can be manifest, even off-shell.

The vector multiplet is a bit more complicated, but in a particular gauge \cite{Ferrara:1974ac}, its components can be taken to be:
\begin{equation}
\label{eq:Vsf}
V= - \theta \sigma^{\mu} \bar{\theta} A_{\mu}^{a}(x) + i \theta^{2} \bar{\theta} \, \lambda^{a \, \dagger}(x) - i \bar{\theta}^2 \theta \lambda^{a}(x) + \frac{\theta^{2} \bar{\theta}^2}{2} D^{a}(x).
\end{equation}
Here, $A_{\mu}$ corresponds to a vector field (e.g., gluon in QCD), $\lambda$ corresponds to its fermionic superpartner (e.g. a gluino) and $D$ corresponds to another non-propagating auxiliary field, which again helps the degrees of freedom balance off-shell. $a$ is the gauge index which runs over the adjoint representation of the gauge group in question.  Another useful object is the (chiral) superfield that packages the gauge field strength
\begin{equation}
\label{eq:FieldStrength}
{\mathcal W}^{a} = -i \lambda^a(y) + \theta D^a (y) - \sigma^{\mu \nu} \theta F^a_{\mu\nu}(y) - \theta^2 \sigma^{\mu} D_{\mu} \lambda^{a} (y).
\end{equation}  
The degrees of freedom here are the same as those enumerated in $V$ above (after our gauge fixing), but both packagings (i.e. $V$ and ${\mathcal W}^{a}$) are useful for constructing parts of the Lagrangian.  It is useful to think of the relationship between $V$ and ${\mathcal W}^{a}$ as the supersymmetrization of the relationship between $A_{\mu}$ and its field strength $F_{\mu \nu}$.

To help us write down supersymmetric theories compactly, it turns out that Grassman integration is quite useful.  For a Grassman variable $\eta$, we have
\begin{equation}
\int d \eta =0 \quad \int \eta \, d \eta =1.
\end{equation}
Thus, an integral of a Grassman variable picks out the coefficient of that variable.  It is useful to introduce the notation $\int d^{2} \theta$ -- which selects out the coefficient of $\theta^{2}$ -- and $\int d^{4} \theta$  -- which selects out the coefficient of $\theta^{2} \bar{\theta}^2$.

We present a few facts without proof (again, refer to, e.g., Refs.~\cite{WessBagger,MartinPrimer} for discussion) that will allow us to construct supersymmetric Lagrangians in an efficient, compact way:
\begin{enumerate}
\item{The product of chiral superfields is itself a chiral superfield.}
\item{The supersymmetric variation of the $F$ component of a chiral superfield is a total derivative, hence when integrated over $d^{4}x$ is invariant.}
\item{ The supersymmetric variation of the $\theta^{2} \bar{\theta^2}$ component of \emph{any} superfield is a total derivative.}
\end{enumerate}


So, working with the basic building blocks of chiral and vector superfields, we can identify a recipe to construct the MSSM. 
\begin{enumerate}
\item{Identify all the left-handed chiral superfields $\Phi_{i}$ that you need based on the particle content of the theory.}\\

\item{We want to generate the kinetic terms for the fields contained in the $\Phi_{i}$.  These arise from ``K\"ahler potential interactions."
Taking 
$$\int{K(\Phi_{i}, \Phi_{i}^{\dagger}) \; d^{4} \theta},$$
selects out a $\theta^{2} \bar{\theta^2}$ term, which is supersymmetric (by (3) above).  The function $K$ is known as the K\"ahler potential.  A minimal K\"ahler potential $\Phi_{i}^{\dagger} \Phi_{i}$ will generate appropriate canonical kinetic terms for a theory of chiral superfields.   If gauge interactions are added, this should be modified as
$$\int{\Phi_{i}^{\dagger} e^{V} \Phi_{i} \; d^{4} \theta}.$$
This generates kinetic terms for the fields in $\Phi$ along with the interactions of these fields with the gauge fields dictated by gauge invariance.  (This requires starting with  Eqs.~(\ref{eq:chiralSF}), (\ref{eq:ydef}), and (\ref{eq:Vsf}) and doing a Taylor expansion).  It also generates interactions between scalars proportional to the gauge couplings once the auxillary field $D$ is eliminated.  These interactions are  required by supersymmetry and are known as $D$ terms.  They will figure prominently in our discussion of the Higgs boson potential.}\\
\item{We can construct another supersymmetric object using the ${\mathcal W}^{a}$ field strength superfield of Eq.~(\ref{eq:FieldStrength}).  For each gauge group write down:
$$\int{{\mathcal W}^{a}(y) {\mathcal W}^{a}(y) \, d^2 \theta}.$$
This is supersymmetric because it is the $\theta^{2}$ component of a chiral superfield. (Remember the product of two chiral superfields is itself a chiral superfield, see (1) above).  This provides the kinetic terms for the gauge fields (and gauginos).
}\\

\item{Finally, choose a \emph{superpotential} $W(\Phi_{i})$.   (Too bad that everything is called $W$.)   This superpotential is a polynomial function of chiral superfields.  Here we use symmetry and renormalizability  as guiding principles.  Notably, supersymmetry imposes that the $W$ is a function of the \emph{chiral} superfields only, and not ``anti-chiral" superfields, $\Phi^{\dagger}$.   We say that the superpotential is a \emph{holomorphic} function. Then (because it is an $F$ term of a chiral superfield) the supersymmetric variation of $\int{W(\Phi_{i}) d^{2} \theta}$ will be a total derivative.  So, the action that results from this construction will be supersymmetric (again, see (1)\&(2) above).}
\
\end{enumerate}
All told, the interactions found by expanding out the expressions in 2., 3. and 4. represent a supersymmetric Lagrangian.

\subsection{How to supersymmetrize the Standard Model}
The above prescription allows us to construct supersymmetric Lagrangians.  Let us now turn to the MSSM itself.
First we identify the superfield content.  There is a vector superfield associated with each gauge group of $SU(3)_C \times SU(2)_L \times U(1)_Y$ --  each of which contains a gaugino, $\lambda_{i}$,  and a gauge field $A^{\mu}$.  There is a chiral superfield for each matter and Higgs field.  It is worth noting that none of the known fermions of the Standard Model can play double duty as as a gaugino  --- none of the SM fermions are in the appropriate adjoint representation. Furthermore, one cannot build a model where the superparter of the neutrino is a Higgs boson.  Although the quantum numbers in this case are basically correct, in these models one would get (much) too large neutrino masses.  With these restrictions in mind, we are led to the particle content of Table~\ref{tab:particleContent}.  Note that the electric charge is given as $Q = T_{3} + Y$.  

The Higgs sector with its two superfields $H_{u}$, $H_{d}$ warrants additional explanation.  In the Standard Model, there is a single complex scalar doublet.  So, we might have thought that  one could promote this scalar to a superfield and be done with it.  However, there are a number of independent arguments that indicate that this is insufficient.  First, the supersymmetrization of the Higgs field introduces a new chiral fermion beyond the Standard Model content (a Higgsino).  The Standard Model content (without this new fermion) is by itself non-anomalous\cite{GrossJackiw}. This means that the presence of a single  Higgsino introduces a gauge anomaly from the triangle diagram containing the Higgsino.  This renders the theory sick at the quantum level.  The introduction of two Higgs multiplets with opposite hypercharge  eliminates this objection.  A separate (also anomaly-based) argument is that adding a single doublet would introduce a global $SU(2)$ anomaly \cite{Witten} -- only theories with even numbers of doublets are non-anomalous.  The addition of a second Higgs multiplet obviates this problem.  

\begin{table}
{\begin{tabular}{@{}cccc@{}} \toprule
Field & SU(3) & SU(2)$_{L}$ & U(1)$_{Y}$ \\ \colrule
$Q$  & 3 & 1 & 1/6 \\
$U^{c}$ & $\bar{3}$ & 1 & -2/3 \\
$D^{c}$ & $\bar{3}$ & 1 & 1/3\\
$L$ & 1& 2 & -1/2 \\
$E^{c}$ & 1 & 1& 1 \\
$H_{u}$ & 1 & 2 & 1/2\\
$H_{d}$ & 1 & 2 &  -1/2 \\
\botrule
\end{tabular}
}
\label{tab:particleContent}
\caption{The Higgs and matter superfields of the minimal supersymmetric standard model.} 
\end{table}

A separate line of argument is that we need to provide masses to all the fermions in the theory.  In the SM, the Higgs boson is in a sense overly efficient.  The Yukawa portion of the SM Lagrangian reads:
\begin{equation}
{\mathcal L} = \lambda_{d} \overline{Q_{L}} \cdot \phi \, d_{R} - \lambda_{u} \epsilon^{ab} \overline{Q_{L}}_{a} \phi^{\dagger}_{b} u_{R} + H.c. 
\end{equation}
The presence of the $\phi^{\dagger}$ presents problems for a naive supersymmetrization.  Recall, to maintain supersymmetry,  the superpotential $W(\Phi^{i})$ can only contain superfields (and not their complex conjugates). Again, the introduction of a second Higgs supermultiplet solves the problem.  We write
\begin{equation}
\label{eqn:WYukawa}
W_{Yukawa} = \lambda_{u} Q H_{u} U^{c} + \lambda_{d} Q H_{d} D^{c} + \lambda_{\ell} L H_{d} E^{c}
\end{equation}
with contraction over relevant indices implied.
The presence of the second Higgs multiplet also allows us to write
\begin{equation}
\label{eq:Wmu}
W_{\mu} = \mu H_{u} H_{d}.
\end{equation}
This term is important because it allows a (Dirac) mass for the fermionic component of the Higgsino multiplet.  
These last two equations, along with the K\"ahler potential and gauge kinetic terms specify the supersymmetric interactions of the MSSM.

\subsection{Interactions and Cancellation of Divergences}
\begin{figure}
\includegraphics[width=0.8\textwidth]{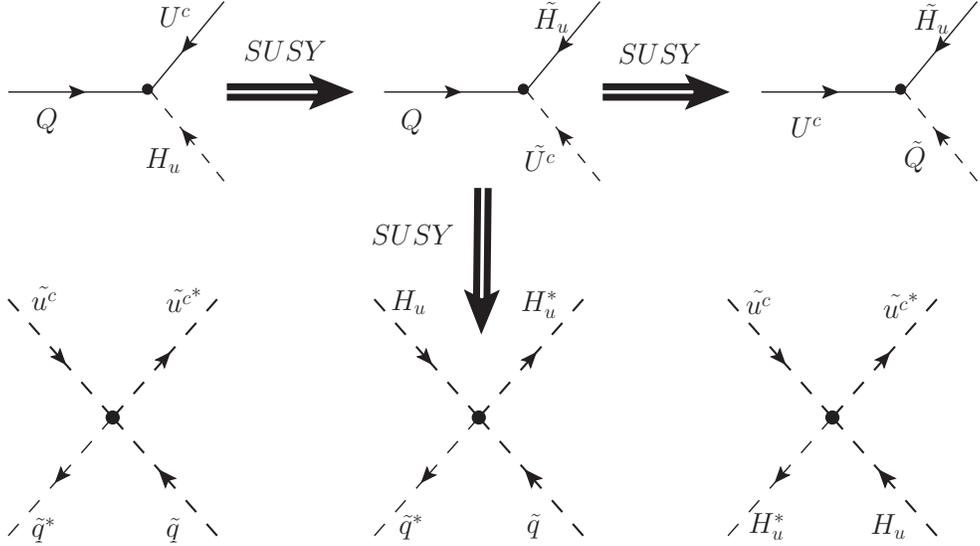}
\caption{Vertices generated by the superpotential interaction of Eq.~(\ref{eqn:testW}).}
\label{fig:vertices}
\end{figure}

Let us write down a superpotential with three superfields, $\{Q, \, U^{c}, \, H_{u}\}$:
\begin{equation}
\label{eqn:testW}
W = \lambda_{u} Q H_{u} U^{c}.
\end{equation}
Assume there is also a minimal K\"ahler potential
\begin{equation}
K = Q^{\dagger} Q + H_{u}^{\dagger} H_{u} + {U^{c}}^{\dagger} U^{c}
\end{equation}
What interactions result from the superpotential of this theory?  Returning to our discussion in
\ref{sec:SUSYBasics}, the prescription is to pick out the $\theta^2$ terms in the expansion of the superfields in Eq.~(\ref{eqn:testW}). One possibility is Yukawa interactions (see Fig.~\ref{fig:vertices}) containing two fermions (which each come with a single power of $\theta$) and a single scalar field (no $\theta$s).  This does not yield just a single
Yukawa interaction -- rather you get three different ones, all with strength $\lambda_u$.  Each superfield gets a turn being the scalar.  These interactions -- and the equality of the couplings --  are forced by the supersymmetry of the theory.  

In addition, this superpotential generates interactions solely involving scalars, the second line of Fig.~\ref{fig:vertices}.  Each of these vertices comes with strength $|\lambda_{u}|^2$.  These interactions arise when Grassman integration picks out two scalar fields (no $\theta$s) and a single auxilary field, $F$ (with its $\theta^2$).  The $F$ can be eliminated using the equation of motion. The Lagrangian contains  an $FF^{\ast}$ term (arising from the K\"ahler potential) and no $F$ term involving derivatives (it is not a propagating field);  so the Euler-Lagrange equations are particularly simple: $\partial {\mathcal L}/ \partial F =0$. This equation sets $F^{\ast}$ equal to a simple function of the scalar fields.  The result is the interactions in   Fig.~\ref{fig:vertices}.

The relationship between the couplings of  the Yukawa interactions in the top line of Fig.~\ref{fig:vertices}
and the quartic scalar interactions in the second line of Fig.~\ref{fig:vertices} is what enforces cancellation of corrections to the Higgs boson mass squared in supersymmetry.  In a supersymmetric theory, a diagram with scalars running around the loop cancels against a diagram with fermions running around the loop because a closed fermion loop comes with an extra (-1).
  
\subsection{Supersymmetry Breaking}
Supersymmetry cannot be an exact symmetry.  If it were, scalars and fermions with identical quantum numbers that would be degenerate.  The absence of a 511 keV charged scalar is a pretty good clue that supersymmetry is at best a broken symmetry.  We will return to how this occurs momentarily, but first let us ask what supersymmetry breaking means for the ultraviolet behavior of this theory.  After all, we imposed supersymmetry in an effort to soften the quadratic divergence of the Higgs boson mass parameters.  Does adding supersymmetry breaking undo all this hard work?  Not necessarily.  In fact, it tells us something about how much supersymmetry breaking we expect in nature.  

\begin{figure}
\includegraphics[width=0.75\textwidth]{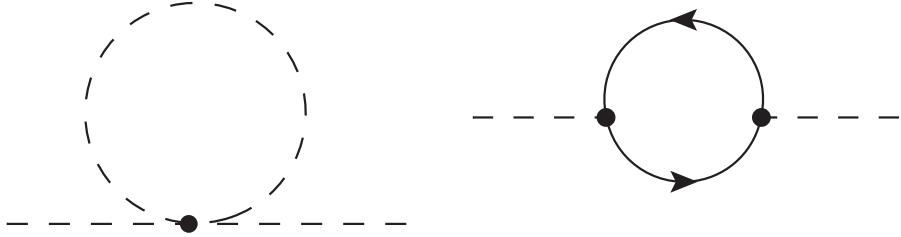}
\caption{Diagrams contributing to $\phi$ self energy in the toy model of Eq.~(\ref{eqn:ToyModel}).}
\label{fig:ToyselfE}
\end{figure}

To understand this, consider a toy model where there is a single superfield $\Phi$ with superpotential
\begin{equation}
\label{eqn:ToyModel}
W_{toy} = \frac{\lambda}{3} \Phi^{3} + \frac{M}{2} \Phi^{2}
\end{equation}
To this supersymmetric theory, we add a scalar mass.  
\begin{equation}
{\mathcal L}_{\cancel{SUSY}} = \mu^{2} |\phi|^{2},
\end{equation}
where $\phi$ is the scalar component of $\Phi$.  The absence of an analogous fermion mass means this breaks supersymmetry. To fully compute the self-energy of the $\phi$ at one-loop there are actually three diagrams.  But to understand how the quadratic divergence is softened, it is sufficient to examine the two diagrams of Fig.~\ref{fig:ToyselfE}.  The one on the left arsies from integrating out the $F$-term.

They yield
\begin{eqnarray}
\rm{Fig. \, 1} &=& \frac{\lambda^{2} }{16 \pi^2} \left( 2 \int{\frac{d^{4}k}{(k^{2} + M^2 +\mu^2)}} - 
\int{\frac{d^{4}k \, Tr[(\not{k} -M)^2]}{(k^{2} +M^2)^2 }}\right) \nonumber \\
&=& \frac{2 \lambda^{2}} {16 \pi^{2}} \int{\frac{d^{4}k \, \mu^2 }{(k^{2} +M^2 +\mu^2) (k^{2} + M^{2})}} \\
&=& \frac{\lambda^{2} \mu^{2}} {8 \pi^{2}} \log{\frac{\Lambda}{M}}, \nonumber 
\end{eqnarray}
where we have cut off the integral at energy $\Lambda$.\footnote{Even for $\Lambda \sim M_{pl}$, this logarithm is manageable, much like in the case of the chiral symmetry of the electron.}  So, while the theory with supersymmetry breaking has a UV divergence, it is only logarithmic in nature.  Furthermore, we note that the coefficient is directly proportional to the size of supersymmetry breaking.   This is the crux of the argument for superpartners  at the weak scale.
 If supersymmetry breaking is ${\mathcal O}$(TeV) or less, then the radiative corrections will not vastly exceed the weak scale.

\subsection{Soft Supersymmetry Breaking Parameters}
\label{sec:SUSYBreaking}
We demonstrated  that supersymmetry breaking scalar masses at the weak scale will not spoil the solution to the hierarchy problem. In fact, there are a variety of supersymmetry breaking terms that do not spoil the UV behavior \cite{Girardello:1981wz}
\begin{itemize}
\item{$m^{2} |\phi|^{2}$: soft scalar masses}
\item{$ B \phi_{i} \phi_{j}$: a mass term between scalars of the same ``holomorphicity".  Due to gauge invariance (and R-symmetry)  this turns out to only be relevant for the combination $H_{u} H_{d}$ }
\item{$A_{ijk} \phi_{i} \phi_{j} \phi_{k}$ : trilinear scalar interactions, or ``$A$-terms"}
\item{$m_{\lambda} \lambda \lambda$:  gaugino masses.}
\end{itemize}
Since the MSSM in the supersymmetric limit is completely controlled by Standard Model parameters, it is variation in the supersymmetry \emph{breaking} parameters that control the variations in phenomenology.  

Where do these Supersymmetry breaking terms come from?   One might have thought that it could be possible to somehow spontaneously break supersymmetry within the MSSM itself.  However, there is a theorem due to Dimopoulos and Georgi \cite{DimopoulosGeorgiThm} that shows that this approach will lead to a squark with mass less than 10 MeV, which is well excluded.   For an elementary review of the proof of this theorem, see Ref.~\cite{CsakiReview}.  

We then are led to a picture where supersymmetry breaking occurs in a so-called hidden-sector quite separate from the MSSM, and this supersymmetry breaking is then subsequently communicated (or ``mediated'') to the visible sector, which is what we call the MSSM.  The details of the mediation mechanism determines how we perceive supersymmetry breaking;  how supersymmetry is broken primordially is less important for the details of the superpartner spectrum.  

One example is so called gravity mediation, wherein supersymmetry is supposed to be broken primordially in another sector, and that supersymmetry breaking is communicated to the visible sector by Planck suppressed operators.  Another example is gauge mediation wherein supersymmetry is broken and communicated to ``messenger" fields that possess Standard Model gauge quantum numbers.  The Standard Model gauge interactions subsequently communicate the supersymmetry breaking to the visible sector.

One consequence of all the allowed supersymmetry breaking terms enumerated above is a proliferation of parameters.    A careful accounting, see, e.g. Ref.~\cite{DineTASI}, yields 105 physical parameters beyond the Standard Model.   These parameters cannot be arbitrary.  If chosen willy-nilly, loop diagrams involving superpartners will lead to dangerous flavor changing neutral currents (FCNCs), yielding unacceptably large contributions to processes like $b \rightarrow s \gamma$, $\mu \rightarrow e \gamma$, and $K \rightarrow \bar{K}^{0}$ oscillations.    This is known as the supersymmetry flavor problem.  If the supersymmetry breaking masses are nearly universal, this substantially mitigates the contributions to these processes.   

In part motivated by this, and in part by a desire for simplicity, a toy spectrum called the constrained minimal supersymmetric standard model (CMSSM) has been widely adopted wherein one posts a single universal gaugino mass $m_{1/2}$, a single universal scalar mass $m_{0}$, and vanishing trilinear terms, all at the GUT scale.   The additional parameters of this ansatz are $\tan \beta \equiv v_{u} / v_{d}$ and the sign of the $\mu$ parameter. Experimental results are often quoted in terms of this parameter space. It should be emphasized that this is a toy model and not the result of an underlying theory.   

Gauge mediation, on the other hand, solves the supersymmetry flavor problem by design -- the only flavor violation will be proportional to that already present in the standard model.  It faces its own challenges, however, including finding a path to an attractive cosmology\cite{Fan:2011ua}.

\subsection{Unification of Couplings}
\label{sec:unification}
With the MSSM structure in place we are equipped to discuss one of the main pieces of ``evidence" for the MSSM: unification \cite{DimopoulosUnify,LangackerUnify}.    Can $g_{Y}, g$ and $g_{3}$ all derive from a single unified gauge group at the high scale?  To quantitatively discuss this question, it is useful to recall the renormalization group equation for a general (not necessarily supersymmetric) gauge theory.  At one loop it is 
\begin{equation}
\frac{dg}{dt} = \frac{b^{1-loop}}{16 \pi^2} g^{3},
\label{eqn:GaugeRGE}
\end{equation}
with $t \equiv \ln{\mu}$ and
\begin{equation}
\label{eqn:oneLoopbGeneral}
b^{1-loop} = -\frac{11}{3} C_{2}(G) + \frac{2}{3} \sum_{f} T_{f}(R) + \frac{1}{3} \sum_{s} T_{s}(R).
\end{equation}
Here, $C_{2}(G)$ is the quadratic Casimir of the gauge group in question.  The quadratic Casimir is defined as $(T^{a})_r^{s} (T^{a})_{s}^t  \equiv C_2(R) \delta_r^t$ for a representation $R$, with the $T^{a}$  in that representation.  For our purposes, we only need $C_2(G)$ the quadratic Casimir for the adjoint representation; $C_2(G)=N$ for SU(N), $N>2$, and $C_2(G) =0$ for a U(1) factor.  This contribution is due to the gauge field self-interactions.  The vanishing of this quantity for a U(1) (which is abelian and lacks gauge self-interactions) is consistent with this interpretation. The $T(R)$ are the so-called Dynkin indices of a  representation $R$, $Tr(T^{a} T^{b})\equiv T(R) \delta^{ab}$.     Note $T(R) = \frac{1}{2}$ for a fundamental of  SU(N), $N>2$, and is equal to the charge squared, $Q^2$, for a U(1) factor.  For an adjoint of SU(N), $N> 2$, we have $T(adj) = N$.
The solution to Eq.~(\ref{eqn:GaugeRGE}) is 
\begin{equation}
\label{eqn:RGESol}
\frac{1} {\alpha_{high}} = \frac{1}{\alpha_{low}} - \frac{b^{1-loop}}{2 \pi} \log{\left(\frac{\mu_{high}}{\mu_{low}} \right) }.
\end{equation}
To check for unification, we can use the experimentally measured values of $\alpha_{i}(M_{Z})$ and ask whether they are consistent with a universal value of $\alpha_{GUT}$.   

Contributions from the SU(3) gauge sector and the three generations of $Q, U^{c}$ and $D^{c}$ yield $b^{SU(3)}_{SM} = -7$.  Similarly, contributions from SU(2) gauge fields, the Higgs doublet,  and the three generations of $Q$ and $L$ yield  $b_{SM}^{SU(2)} = -19/6$.  Finally, carefully accounting for multiplicity of generations, doublets, and color, gives $b^{Y}_{SM} = 41/6$.  

For supersymmetric theories, Eq.~(\ref{eqn:oneLoopbGeneral}) simplifies.
\begin{equation}
b^{1-loop}_{SUSY} = - 3 C_{2} (G) + \sum_{r, matter} T_{r},
\end{equation}
where we have used the fact that there is a fermionic representation in the adjoint (gauginos) and that all matter representations have both fermions and scalars.  With MSSM field content, it is straightforward to compute  $b_{SU(3)}^{MSSM} = -3$, $b_{Y}^{MSSM} = 10$, and $b_{SU(2)}^{MSSM} = 1$.  If we assume the superpartners all lie at the weak scale (more on this below), we can use these supersymmetric beta functions to evolve the measured values of gauge couplings at the weak scale up to high scales and check for unification.

Before checking unification, we must deal with a subtlety: $b^{Y}$ and $g^{Y}$ are not properly normalized to be embedded in a grand unified theory. To ensure proper normalization, we want Tr(generator$^{2}$) to be equal for any gauge generator in the unified group, where the trace runs over all particles in a representation.  Consider a $\bar{5}$ of SU(5) 
\begin{equation}
\bar{5} = \left(\begin{array}{c}  
d^{c}\\
d^{c}\\
d^{c}\\
\left(\begin{array}{c}\nu \\
e^{-}\end{array} \right)
\end{array}
\right).
\end{equation}
 For color generators $SU(3)_C$, we have 
$Tr(T^{a} T^{b})= T(R) \delta^{ab}$, with $T(R) = \frac{1}{2}$ coming from the the (anti-)fundamental down-quark.  On the other hand, for hypercharge, $Tr(Y^2) = 3\times (1/3)^2 + 2 (1/2)^2 = 5/6$.  Since $5/6  \neq 1/2$ we need a new normalization for our $U(1)$.  We cannot arbitrarily change the physical combination of couplings and quantum numbers, so we must take
\begin{equation}
g_{Y} Y = g_{1} Y^{\prime},
\end{equation}
such that
\begin{equation}
Y^{\prime} = \sqrt{\frac{3}{5}} Y  \hspace{.5in} g_{1} = \sqrt{\frac{5}{3}} g_{Y}.
\end{equation}

We are now (finally) in a position to check unification.  At one loop, we find (imposing a universal GUT coupling at the high scale)  Eq.~(\ref{eqn:RGESol}) implies 
\begin{equation}
\frac{b_{3} - b_{2}}{b_2 -b_{1}} = \left.\left(\frac{\alpha_3^{-1} -  \alpha_2^{-1}}{\alpha_2^{-1} - \alpha_1^{-1}}\right) \right|_{M_{Z}}.
\label{eqn:unificationFOM}
\end{equation}
Using experimentally determined values \cite{PDG} of the couplings on the RHS at $M_{Z}$, we find $RHS=.718 \pm 0.003$.   On the other hand, substituting our beta functions above for the LHS we find $LHS_{SM} = .528$ and $LHS_{MSSM}= .714$.  The Standard Model does not appear to unify.  The MSSM seems plausibly consistent with unification.  We did not take into account (important) two-loop effects, or threshold effects (from both the weak scale and GUT scale) arising from mass splittings within grand unified multiplets.  Finally, we have assumed all the superpartners lie right at $M_{Z}$, and apparently, they do not.  So, the precise agreement demonstrated here for  the MSSM should be somewhat taken with a grain of salt, but it is reasonable to say that unification plausibly occurs in the MSSM once percent level threshold corrections are taken into account.


This unification of couplings is perhaps the single-most compelling piece of evidence (aside from the hierarchy problem itself) for finding supersymmetry at the weak scale.  It is possible that it is a coincidence, but it is striking.  Without this evidence, it is unlikely that supersymmetry would have received quite so much attention from the theoretical community.

\section{Basics of Supersymmetry for Colliders}

The goal of this section is to give a broad brush appreciation of the kinds of signatures that the MSSM can produce and are searched for at colliders.  

\subsection{Where is SUSY?}
Building on the arguments of Sec. \ref{sec:WhySUSY}, we can ask a bit more precisely where we expect supersymmetry to show up.  Consider the one-loop correction to the up-type Higgs $(mass)^2$ parameter.  Interactions proportional to the top quark Yukawa coupling (see Fig.~\ref{fig:TopLoop})  would be quadratically divergent.   However, when supplemented by diagrams involving the stop squark, they are softened to:
\begin{equation}
\label{eqn:stoploopcorr}
\Delta m_{H_{u}}^{2} = - \frac{3 \lambda_t^{2}} {8 \pi^{2}} \left( \tilde{m}^{2}_{Q_3} + \tilde{m}^{2}_{u_3}   + |A_{t}|^2 \right)  \log \left( \frac{ \Lambda}{m_{\tilde{t}}} \right),
\end{equation}
with $\tilde{m}$ representing soft scalar masses, and $A_{t}$ a trilinear coupling for the top squarks.  The soft supersymmetry breaking parameters appear explicitly (just as in our toy model) out in front -- the correction to the Higgs mass is quadratically sensitive to these.  Furthermore, there is a logarithmic enhancement, cutoff at some scale $\Lambda$.  This is the scale at which the soft masses explicitly ``become soft".   The value of $\Lambda$ depends on the supersymmetry breaking and mediation mechanism and can lie anywhere between 100 TeV and the Planck scale.\footnote{In the case of gauge mediation, $\Lambda$ corresponds to the messenger masses.   For gravity mediation, the Planck scale would be appropriate; so the logarithm largely cancels the loop factor that appears out in front.}  All told, this equation suggests that the third generation superpartners should be quite close to a TeV.  

Also, somewhat less directly (but importantly), there is a correction to the stop mass coming from the gluino
\begin{equation}
\Delta m_{\tilde{t}}^2 = \frac{8 \alpha_{s} } {3 \pi} |M_{3}|^2 \log{\left(\frac{\Lambda}{m_{\tilde{t}}} \right)}.
\end{equation}
A too-large gluino mass will feed into the stop mass, and then quickly feed into the Higgs mass parameter via Eq.~(\ref{eqn:stoploopcorr}).  The combination of these facts argue that a natural supersymmetric theory wants both gluinos and stops to be ``somewhat light".  It is less clear where all the other superpartners must lie  (though  naturalness considerations also suggest light Higgsinos, more on this below). It might be that all superpartners are fairly close in mass -- naturalness considerations alone, however, would allow the superpartners of the first two generations to be quite a bit heavier.    

So the above arguments suggest that stops and gluinos (at least) should be produced at the LHC if the theory is to avoid fine-tuning.  As the limits on gluinos and stops increase, this means that the possibility of a natural theory of supersymmetry is becoming more and more squeezed.  But before we are in a position to talk about how to see gluinos, stops, and other superpartners, we need to understand how they behave in the detector.  Some will immediately decay, but -- at least in the most common version of the  MSSM-- the lightest will not.  To understand why this is so requires an examination of a new kind of symmetry possible in supersymmetric theories: the ``R-symmetry".

\subsection{R-symmetries and the fate of the Lightest Superpartner}
\label{sec:Rsym}
 Under an R-symmetry, the Grassman $\theta$ parameters that enter the superfield expansion transform:
\begin{eqnarray}
\theta \rightarrow e^{i \alpha} \theta \; &;& \; R[\theta] =1\\
\bar{\theta} \rightarrow e^{-i \alpha} \bar{\theta}  \; &;& \; R[\bar{\theta}] =-1
\end{eqnarray}
An immediate consequence is that  R-symmetry treats different components of a super-multiplet differently, see Table \ref{tab:Rcharges}. 
The $R$ charge of the superfield is equal to the  combined $R$-charge of the component fields with the R-charges of the $\theta$'s that multiply the component fields in the superfield expansion. 

\begin{table}
{\begin{tabular}{@{}cc@{}} \toprule
Field & R  \\
\hline
Vector Superfield $V$  & 0\\
gauge field: $v_{\mu}$ & 0  \\
gaugino: $\lambda$ & 1  \\
auxiliary D& 0\\
\botrule
\end{tabular}
\quad \quad \quad
\begin{tabular}{@{}cc@{}} \toprule
Field & R  \\
\hline
Chiral Superfield $\Phi$  & q\\
scalar field: A & q  \\
fermion: $\psi$ & q-1  \\
auxiliary F&  q-2\\
\botrule
\end{tabular}
}
\label{tab:Rcharges}
\caption{R Charges of Superfields and components} 
\end{table}

The K\"ahler potential term $\int d^{4} \theta \, Q^{\dagger} e^{V} Q$ is invariant upon imposing $R[V]=0$.    What condition does $R$-invariance impose on the superpotential?  Since $\int d^{2} \theta$ effectively strips away two $\theta$'s, we can write $R[d^{2} \theta] = -2$, and if $R[W] =2$ the Lagrangian will be invariant, see Sec.~\ref{sec:SUSYBasics}.

Now suppose we write down a superpotential that respects this symmetry.  An example is the superpotential of the MSSM
\begin{equation}
\label{eqn:Rinv}
W = \lambda_{u} Q U^{c} H_{u} + \lambda_{d} Q D^{c} H_{d}   + \lambda_{\ell} L E^{c} H_{d} + \mu H_{u} H_{d},
\end{equation}
with $R_{Q} = R_L =R_{H_{u}} = R_{H_{d}} = 1$, $R_{U^{c}} = R_{D^{c}}=0$.

What does soft  supersymmetry breaking do to this symmetry?  We can use the spurion language (discussed in detail in J.~Thaler's lectures at this school \cite{ThalerLec})  to encapsulate the effects of supersymmetry breaking.  We parameterize the effects of supersymmetry breaking by giving a vacuum expectation value to the $F$ term of a spurion field $Z$:
\begin{equation}
Z = m \theta^{2}
\end{equation}
Effectively this $Z$ field has an $R$-charge of 2 due to the presence of the two $\theta$s in its expansion.  The spurion field is a shorthand device that allows tracking of the effects of supersymmetry breaking.  The actual dynamics that gives rise to the supersymmetry breaking may be (much) more complicated than a single field.

This spurion allows us to write down scalar soft masses, for example, for squarks:
\begin{equation}
 \int d^{4} \theta \, Q^{\dagger} Q  Z^{\dagger} Z \rightarrow m^{2} \tilde{q}^{\ast} \tilde{q}.
 \end{equation}
The $\theta^2$ and $\bar{\theta}^2$ from the $Z$ and $Z^{\dagger}$ saturate the $d^{4} \theta$ integral, and only the lowest components of the $Q$ superfields are relevant.
This term is R-invariant; the non-zero $R$ charge of $Z$ is compensated by the opposite $R$-charge of $Z^{\dagger}$.  So, scalar soft masses do not violate $R$.  On the other hand, the scalar trilinear couplings $A_{ijk} \phi_{i} \phi_{j} \phi_{k}$ are given as, e.g., 
 \begin{equation}
 \int d^{2} \theta Q U^{c} H_{u} Z \rightarrow m \tilde{q} \tilde{u^{c}} h.
 \end{equation}
 Note that the superpotential of Eq.~(\ref{eqn:Rinv}) absent $Z$ was R-invariant.  So, this means the inclusion of this term breaks $R$ by two-units -- the two units present in $Z$.  Similarly, gaugino masses can be written down with the help of the spurion and the field strength chiral superfield ${\mathcal W}$
 \begin{equation}
 \int d^{2} \theta Z {\mathcal W} {\mathcal W} = m \lambda \lambda.
\end{equation}
This will also break $R$ by two units.  So soft-terms break the $U(1)_{R}$ that we had present in our original theory.  But the symmetry is not broken entirely  --   since the soft supersymmetry breaking terms only break $R$ by two units, the $R$-symmetry  is broken down to a residual parity, which we can denote as $R_{p}$.  This residual parity will have important implications.

It turns out that the precise $R_{p}$ as derived from the charge assignment  in Eq.~(\ref{eqn:Rinv}) is not so useful  because the Higgs fields acquire vacuum expectation values, and so this naive parity is spontaneously broken.  A more useful symmetry that is preserved is one where we have $\theta  \rightarrow - \theta$, $H_{u,d} \rightarrow H_{u,d}$, and $(Q,U^{c},D^{c},L,E^{c}) \rightarrow -(Q,U^{c},D^{c},L,E^{c})$.  This parity symmetry,  a combination of fermion number and matter parity, is the $R$-parity of the MSSM.  With this choice, all Standard Model fields (including the Higgs scalars) are even, and all superpartners are odd. 

Conservation of this symmetry is crucial for determining the collider phenomenology.  In particular, it enforces 
\begin{itemize}
\item{Superpartners must be produced in pairs.  We start in a R-parity even state, and end in an R-parity even state.}
\item{The lightest superpartner is stable, as it has no R-odd final state to decay into.  This means that super partners will exit the detector as missing energy.}
\end{itemize}

This symmetry forbids terms that would otherwise be allowed by the combination of gauge symmetry and renormalizability:
\begin{eqnarray}
W_{\Delta L} &=& \mu^{\prime} L H_{u} + \lambda_{ijk} L_{i} L_{j} E^{c}_{k} + \lambda^{\prime} Q_{i} D_{j} L_{k} \\
W_{\Delta B} &=& \lambda^{\prime \prime} U_{i}^{c} D_{j}^{c} D_{k}^{c}.
\end{eqnarray}
(Alternately, the presence of all these terms preclude an $R$-parity).  These terms,  if large, are a phenomenological disaster.  They can mediate large flavor changing neutral currents as well as rapid proton decay. Note that the potential existence of these interactions can be viewed as a step backwards from the Standard Model.  There, renormalizability and gauge invariance are sufficient to strongly suppress these kinds of processes.  For example, if the Standard Model were valid up to the Planck scale, one might guess that the leading contribution to proton decay would be dimension-six operators suppressed by two powers of $M_{pl}$.    One attitude frequently taken is that successful supersymmetric phenomenology motivates the imposition of an $R$-symmetry, and that this symmetry, amongst other things, gives rise to an excellent Dark Matter candidate (see lectures by R.~Essig and S.~Profumo \cite{ProfumoLec} at this school).

\subsection{Who are we producing anyway?  Cast of Characters: Mass eigenbasis}
After electroweak (and super-) symmetry breaking, the mass eigenstates do not align with the superfields.  Fields with identical electric charge but different electroweak quantum numbers are expected to mix.  In particular, amongst neutral fields, we expect mixing between the superpartners of the Higgs fields (known as Higgsinos) and the superpartners of the neutral $SU(2)$ gauge boson (the wino, $\tilde{w^{0}}$) and the $U(1)_Y$ gauge boson, known as the bino ($\tilde{B}^0$). These fields are collectively known as neutralinos. All told, have the symmetric matrix (written in the basis $\{ \tilde{B}^{0}, \tilde{w^{0}}, \tilde{H}_{d}, \tilde{H}_{u} \}$):
 \begin{equation}
 M_{neutralino}= \left(\begin{array}{cccc}
 M_{1} & 0         & -\frac{1}{2} g_{Y} v_{d} &  -\frac{1}{2} g_{Y} v_{u} \\
 0         & M_{2} & -\frac{1}{2} g v_{d} &  -\frac{1}{2} g v_{u} \\
  -\frac{1}{2} g_{Y} v_{d}   & -\frac{1}{2} g v_{d}    & 0       &  -\mu \\
 -\frac{1}{2} g_{Y} v_{u}    &-\frac{1}{2} g v_{u}   & -\mu    &  0   
 \end{array}
 \right).
 \end{equation}
The $M_{1}$ and $M_{2}$ are Majorana gaugino mass terms that arise from supersymmetry breaking, see Sec.~\ref{sec:SUSYBreaking}.  The $\mu$ terms arise directly from the superpotential Eq.~(\ref{eq:Wmu}).  The terms in the upper-right hand block arise from terms in the Lagrangian like ${\mathcal L} \ni g^{\prime} \tilde{B} \tilde{H_{u}}  \langle H_{u} \rangle$.  Their presence is required by the  combination of gauge interactions and supersymmetry.  Note that these interactions are just Yukawa couplings: they have two fermions ($\tilde{B}$ and $\tilde{H_{u}}$ in our example) that are married together once the Higgs field $H_u$ takes on its expectation value.  Because of the additional (super)symmetry, this \emph{a priori} independent Yukawa coupling is related to a gauge coupling.  This mass matrix can be diagonalized by an orthogonal matrix. The mass eigenstates are  $\chi^{0}_{1,2,3,4}$ with  $M^{diag}_{neutralino} = N^{T} M_{neutralino} N$.

A similar story obtains for the charged sector.  The mass matrix for the charginos is
\begin{equation}
M_{chargino}  = \left(\begin{array}{cc}
 M_{2} & \sqrt{2} \sin{\beta} M_{W}          \\
\sqrt{2} \cos{\beta} M_{W}       & \mu 
 \end{array}
 \right), 
\end{equation}
where we have traded the $M_{W}$ for the product of the Higgs vacuum expectation value (vev) and gauge coupling. Here the columns are labelled by $\tilde{w}^+$ and $\tilde{H}_{u}^+$and rows are labelled by $\tilde{w}^-$ and $\tilde{H}_{d}^-$.  The diagonal elements are present due to supersymmetry breaking gaugino mass $M_{2}$ and supersymmetric Higgsino mass $\mu$.  The off-diagonal terms are again due to the Yukawa couplings that arise upon supersymmetrization of gauge interactions.  The end result is that the wino and Higgsinos marry off to form a pair of Dirac fermions, denoted  $\chi^{\pm}_{1,2}$.  Because this  $2 \times 2$ matrix is not symmetric,  two different rotation matrices to do the  singular-value decomposition of this matrix, $M_{diag}^{chargino} = U M_{chargino} V^{\dagger}$: 
\begin{equation}
\left( \begin{array}{c}
\chi_{1}^{+} \\
\chi_{2}^{+}
\end{array}
\right)
= V \left( \begin{array}{c}
\tilde{w}^{+} \\
\tilde{H_{u}}^{+}
\end{array}
\right),
 \quad
\left( \begin{array}{c}
\chi_{1}^{-} \\
\chi_{2}^{-}
\end{array}
\right)
= U \left( \begin{array}{c}
\tilde{w}^{-} \\
\tilde{H_{d}}^{-}
\end{array}
\right).
\end{equation}


\subsection{Production Modes}
\begin{figure}
\includegraphics[width=\textwidth]{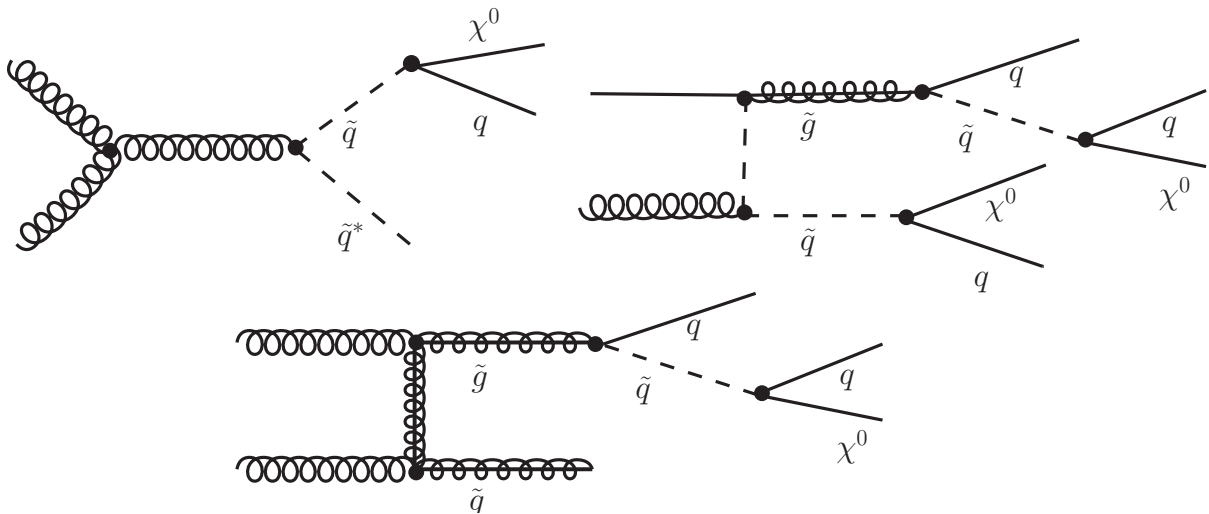}
\caption{Production of colored superpartners.  Clockwise from upper-left are representative diagrams for squark pair production, squark-gluino associated production and gluino pair production.  Their decays are shown to quarks and the lightest supersymmetric particle, here assumed to be the neutralino.  The result is a final state with jets and missing energy.}
\label{fig:Coloredproduction}
\end{figure}

With the cast of characters in place, let us examine the production modes that are most relevant for the LHC.  Because the LHC is a hadron collider, there are plenty of gluons and quarks in the initial parton distribution functions.  The result is that if colored states are kinematically accessible, they are likely to dominate production.  Pair production of colored particles is shown in Fig.~\ref{fig:Coloredproduction}.  To set the scale, production cross sections are shown in Table~\ref{tab:productionXsec}.  Its apparent that the processes that are being searched for make up a tiny fraction of the total event rate at the LHC.  So, to identify these rare processes, care must be taken to cut away the dominant Standard Model backgrounds.  Details of these experimental searches are discussed elsewhere at this school.

\begin{table}
{
\begin{tabular}{@{}ccc@{}} \toprule
Process & LHC8 & LHC 14  \\
Total &   .1b & .1 b\\
$b \bar{b}$ ($p_T> 30$ GeV) &  $.3 \; \mu$b    & $1 \;\mu$b \\
$t\bar{t}$&  200 pb & 800 pb \\
$gg \rightarrow h$ & 15 pb & 50 pb \\
gluino ($m_{\tilde{g}}$ = 500 GeV) & 4 pb & 30 pb \\
gluino ($m_{\tilde{g}}$ = 750 GeV) &   200 fb    & 3 pb \\
gluino ($m_{\tilde{g}}$ = 1 TeV) &   20 fb       &  400 fb \\
\botrule
\end{tabular}
}
\label{tab:productionXsec}
\caption{Order of Magnitude Production Cross Sections at LHC, data taken from Ref.~\cite{HanTASI, GluinoProduction}.}
\end{table}

\subsection{Classic SUSY Signals}
We are finally ready to discuss a few of the classic signatures of supersymmetry.   As discussed above, because the LHC is a hadron collider, it is likely that the dominant production of superpartners will be colored ones, as shown in Fig.~\ref{fig:Coloredproduction}.  The colored particles will then decay down to the lightest supersymmetric particle, which is stable, as described in Sec.~\ref{sec:Rsym}.  It leaves the detector, carrying energy with it.  This ``missing energy'' signature is a hallmark of supersymmetry searches. After quarks hadronize to form jets, the final state will be jets and missing energy.  This search is an excellent all-purpose search for supersymmetry, and in much of the parameter space of simplified models (e.g., the CMSSM described above), it provides the strongest bounds.

\begin{figure}
\begin{center}
\includegraphics[width=0.65\textwidth]{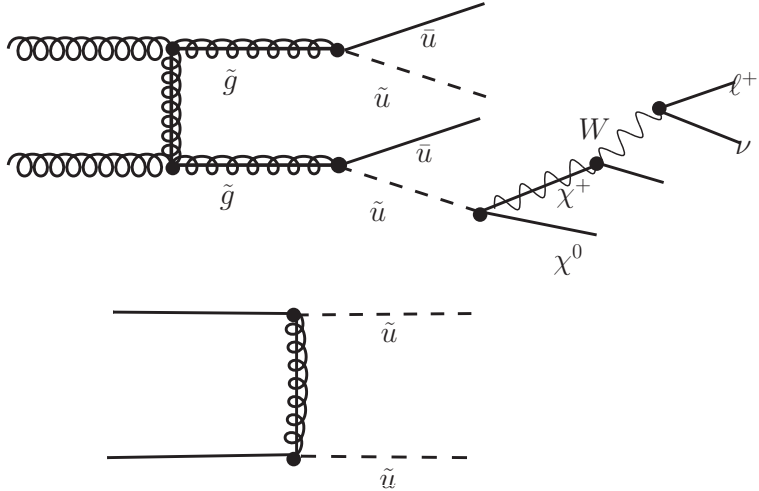}
\caption{Processes leading to same sign leptons and missing energy.  At top, gluino pair production can lead to production of same signed quarks in their decay.  At bottom, direct production of same sign squarks is shown.  Subsequent decay of the squarks can lead to same sign leptons.}  
\label{fig:SSDL}
\end{center}
\end{figure}

Another classic signature is the production of like-sign dileptons.  In the Standard Model, most pair production processes (e.g. $t \bar{t}$, $WW$) and decays (e.g. $Z$ decay) typically lead to opposite sign leptons.  In supersymmetry things can be different.  At the LHC, two initial state $u$-quarks can form two $u$-squarks via the $t$-channel exchange of a gluino.  The two $u$ squarks can then ``cascade'' decay down to the LSP.  Because the squarks have the same sign, it is possible that like-sign $W$'s (and hence leptons) can be produced (see Fig.~\ref{fig:SSDL}).  Two squarks of the same sign can also be produced as a result of gluino pair production (see Fig.~\ref{fig:SSDL}).  The reason is that the gluino does not carry quark number.   Dominant standard model backgrounds include $t \bar{t} + W/Z$, $W^{\pm} W^{\pm} qq, W^{\pm} Z$.

\begin{figure}
\includegraphics[width=.75\textwidth]{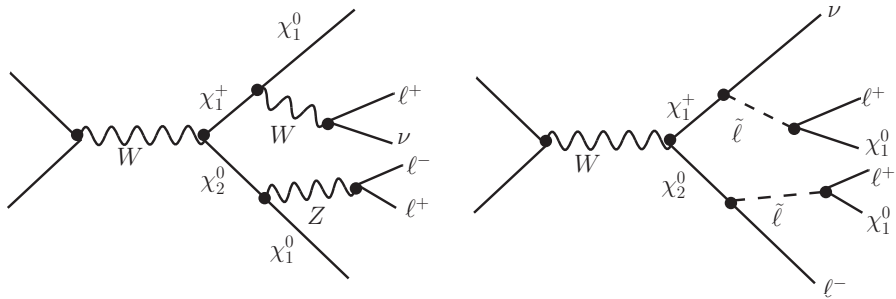}
\caption{Production of charginos and neutralinos leading to trilepton signals.  Decays may proceed via gauge bosons or potentially via sleptons.  The result is a final state with three leptons and missing energy.}
\label{fig:trilepton}
\end{figure}

The last classic signal we will mention is the trilepton search.  This can be produced via the direct production of charginos and neutralinos (Fig.~\ref{fig:trilepton}). The leptons can arise either through the decays of gauge bosons, or if the sleptons are relatively light.  

As of yet, not excesses have been seen in any of these channels.  The latest supersymmetry searches at ATLAS and CMS can be found at Ref.~\cite{ATLASSUSY}, \cite{CMSSUSY}.   The purpose of these notes is not to review the present searches in any detail (see lectures by V.~Sanz at this school for some related case studies).  Depending on the details of the superpartner spectrum,  current limits can exclude superpartners up to masses of nearly a TeV.  It is possible to soften these limits somewhat in cases of specific designer spectra.  Nevertheless, given our desire to have superpartners near the weak scale to tame the quadratic divergence in the Higgs boson mass parameter, it is fair to say things are becoming somewhat uncomfortable for the most natural instantiations of supersymmetry.   

 \subsection{Not so Classic Signals}
 Our guiding principle in our discussion of collider phenomenology was the existence of a R-parity.  While there is strong motivation for this symmetry, there is no guarantee of its existence.  For example, it is possible to avoid proton decay without an imposing the full R-symmetry.  For example, conservation of baryon number (to forbid $W_{\Delta B}$) is  enough to ensure proton stability.  Actually, the absence of either  $W_{\Delta L}$ or $W_{\Delta B}$ is enough\footnote{This conclusion holds as long as the gravitino is not too light\cite{ChoiProtonDecay}.}.  Recently, there has been a resurgence in activity in examining models with such a reduced symmetry -- in part because the collider phenomenology differs dramatically from that described above, and it allows a weakening of the constraints on the supersymmetric parameter space.  Typically $R$ parity violation is a  small perturbation, so that all heavy superpartners that are produced will decay down to the lightest superpartner, which (since it has no other choice) uses the $R$-parity violating couplings to decay.  This last decay degrades the missing energy signal. Instead, typically, events have a large multiplicity.  For example, if the superpotential contains the $U^{c} D^{c} D^c$ operator, the lightest neutralino might decay to three quarks.


Another direction of interest is to examine compressed spectra \cite{MartinCompressed,LeCompte:2011fh}.  The missing energy observed at colliders depends largely on mass splittings, and not as directly on the overall mass scale.  Thus, if the superpartner spectrum is compressed, the missing energy signals will be degraded, this somewhat softens the limits on the superpartners.

\subsection{Split Supersymmetry}
\label{sec:split}
We spent a long time arguing for a scale of supersymmtery breaking $\tilde{m} \approx M_{W}$ on the basis of naturalness.  However, it should be noted that there is an example where the idea of naturalness fails spectacularly.  If one were to guess the size of the cosmological constant (CC)  -- completely unprotected by any symmetry -- the guess would be  $\Lambda_{CC}^{4} \sim (\rm{cutoff})^4$.  If the Standard Model is valid up to the Planck scale, this estimate is off by 120 orders of magnitude: the true value of the CC is $(\rm{meV})^4$.  Even if there were supersymmetry right at the TeV scale (which would cutoff contributions to the CC above this scale),  we would be wrong by sixty(!) orders of magnitude.  What are we to make of this? Perhaps the CC, which after all depends (in part) on a quantum theory of gravity, differs fundamentally from the Higgs mass parameter, whose fine-tuning we were so worried about?    Maybe one day, with a fuller understanding of String Theory, we will understand why the CC is so small.  Of course, this need not be so, in which case, the spectacular failure of naturalness might give one pause about the likelihood of observing supersymmetry. And even if this is so, we might worry that the solution to the CC problem does not commute with the hierarchy problem.  If it does not, we might worry that a universe that has the right cosmological constant might look fine-tuned in the Higgs sector \cite{SSDimopoulos, GiudiceSplit}.

But if the Higgs sector is fine-tuned is there \emph{any} reason to talk about supersymmetry at all?  Suppose, e.g., that string theory ``likes" supersymmetry -- perhaps it is necessary for the ultimate stability of the theory -- but the scale of supersymmetry breaking has nothing to do with the hierarchy problem (or at best solves it incompletely).  Then, at what scale should we see superpartners?  Absent the fine-tuning motivation, it turns out there is not really any scale strongly preferred for the scalar masses\footnote{Except perhaps by some indirect cosmological arguments, see, e.g., Ref.~\cite{PierceWacker}.  Also, the 125 GeV Higgs mass may favor scalar mass that are not \emph{too} heavy.}.    The fermionic superpartners are another story. They are responsible  for the improvement in gauge coupling unification with respect to the Standard Model.  The reason is that the new scalar superpartners (all the squarks and sleptons) come in complete SU(5) multiplets.  Carefully examining our discussion in \ref{sec:unification}, we see that complete GUT multiplets do not contribute to the LHS of Eq.~(\ref{eqn:unificationFOM}).     So, if we are to maintain successful unification of couplings these fermions should be ``somewhat close" to the TeV scale.  Unfortunately, this unification argument is not strong enough to say they should live below 1 TeV, and the difference between 1 TeV and 10 TeV is everything when it comes to the LHC.  An independent argument for relatively light gauginos is that they make up an excellent Dark Matter candidate. Again, a successful cosmological history typically indicates gaugino masses near a TeV, but the argument is not sharp enough to say exactly where they would lie.  For example, a pure wino (with very small admixture of other neutralinos) can give the right thermal relic abundance of Dark Matter if its mass is somewhat above 2 TeV, but a proper admixture of bino and Higgsino could be a couple hundred GeV.  

The end result of this line of reasoning is a ``split spectrum" wherein the scalar can be (quite heavy), but fermions (gauginos and Higgsinos) are near the weak scale.  How large the splitting can be is model dependent.  One especially simple scenario involves a splitting of approximately one-loop factor between the scalars and fermions, a scenario discussed in Ref.~\cite{OriginalAMSB, WellsPeV}, and recently revisited in Ref.~\cite{DimopoulosMinisplit, HallNomuraSpread}.  A similar spectrum was also motivated from a ``top-down" approach from a string compactification \cite{KaneG2}.

How can one test this scenario?  One can look for the Dark Matter in direct \cite{GiudiceSplit, PierceSSDM} or indirect \cite{SSIndirect1, SSIndirect2, Alves:2011ug} detection experiments.    Direct production of charginos and neutralinos at the LHC may also be possible.  Perhaps one of the most interesting searches is for gluinos.  Because the squarks are potentially heavy, the lifetime of the gluinos can in principle be quite long.  If the squarks are sufficiently heavy, the gluinos transit macroscopic distances in the detector before decaying.  In fact, it is possible the gluinos hadronize, eventually coming to a stop \cite{Arvanitaki:2005nq} in the detector prior to decaying.  Decays out of synch with the beam (perhaps even occurring when the LHC is off!) would be a striking signature indeed.  Thus far, no such decays of long-lived hadrons have been observed, and limits of near a TeV have been placed, see Ref.~\cite{ATLAS:2012al,CMSStoppedGluino,CMS:2012yg,Chatrchyan:2012sp,Aad:2012zn}.  However, it should be noted that of parameter space of theories with the heaviest scalars (and most striking gluino signatures) can give too heavy Higgs boson masses.  It may be that gluino lifetimes are only long enough to allow slightly displaced vertices (if at all).


\section{Electroweak symmetry breaking in the MSSM}
But the absence of supersymmetric signatures is not the most exciting data we have received from the LHC to date.  That honor falls on the discovery of a 125 GeV Higgs boson.  What are the  implications this discovery  for supersymmetry? Before we can answer this, we need to understand a bit about the Higgs potential in supersymmetry.  As alluded to above, the MSSM is a two Higgs doublet model (2HDM).  
For the two Higgs doublets of hypercharge $Y= \pm 1/2$, we have scalar multiplets
\begin{equation}
\left(
\begin{array}{c}
 H_{u}^+ \\\
H_u^0
\end{array}
\right),
\hspace{1in}
\left(
\begin{array}{c}
 H_{d}^0 \\\
H_d^-
\end{array}
\right).
\end{equation}
However,  supersymmetry restricts the form of the potential with respect to a general 2HDM.  In particular, some quartics are not  allowed, and others are fixed to values related to the gauge couplings.  This is to be expected -- any time you impose a symmetry on a theory, there are relationships that you would not otherwise would have expected.  After imposing these constraints, we find 
\begin{eqnarray}
V(H_{u}, H_{d}) = 
m_{2}^{2} |H_{u}|^{2} + 
m_{1}^2 |H_{d}|^2 - \left(\mu B (H_{u} \cdot H_{d}) + c.c.\right) \nonumber \\
+ \frac{g^{2}}{2} ( H_{d}^{\ast} \frac{\tau}{2} H_{d}  + H_{u}^{\ast} \frac{\tau}{2} H_{u} )^{2} +  \frac{g_{Y}^{2}}{2} ( \frac{1}{2} |H_{u}|^{2}  -\frac{1}{2}|H_{d}|^{2} )^{2}.
\end{eqnarray}
Note, $m_{1,2}^2 \equiv \mu^2 + m_{H_{d,u}}^2$ with $m_{H_{d,u}}^2$ the relevant soft supersymmetry breaking scalar $(mass)^2$ parameters.  The $\tau$ correspond to the Pauli matrices.  

There are a variety of different contributions to this potential.  Some are supersymmetry preserving contributions arising from the superpotential $(\propto \mu^{2})$, some arise from supersymmetry breaking, and others (those proportional to gauge couplings) derive from the $D$-terms.   The $(B \mu H_{u} H_{d}+ h.c.)$ term breaks not only supersymmetry (again softly), but an additional global symmetry that takes $H_{u} \rightarrow e^{i \alpha} H_{u}, H_{d} \rightarrow e^{i \alpha} H_{d}$.  So, we expect there to be an additional massless Goldstone boson in the limit where this term vanishes. 

After doing some rearranging, we can rewrite the $D$-term contributions as:
\begin{equation}
V_{D} = \frac{\bar{g}^{2}}{8} (|H_{u}|^{2} - |H_{d}|^{2})^2 + g^{2} |H_{u}^{\dagger} H_d|^2
\end{equation}
where we have defined $\bar{g^2} \equiv g^{2} + g_{Y}^2$.  It is worthwhile reemphasizing that these quartic coupling of the Higgs fields are related to the gauge couplings because of supersymmetry.

Let us explore the minimization of this potential.  Using an $SU(2)_{L}$ gauge transformation, we can set, say, $H_{u}^+ =0$.  That is, we can rotate the vev into one component of the doublet.  It is a straightforward exercise to check that 

\begin{equation}
\frac{\partial V}{\partial H^{+}_{u}} =0 \rm{\; at \; } H_{u}^{+}=0 \Rightarrow H_{d}^- =0.
\end{equation}
This is good news, as it means that electric charge will not be broken at the minimum of the potential (no charged component gets a vacuum expectation value).  Since only neutral components will get vevs, we can rewrite the potential as a function of $v_{u}$ and $v_{d}$, where
\begin{equation}
H_{u} = \left( 
\begin{array}{c}
0 \\
v_{u}
\end{array} 
\right)
\quad
H_{d} = \left( \begin{array}{c}
v_{d}\\
0
\end{array} 
\right).
\end{equation}
In principle these $v_{u}$ and $v_{d}$ could be complex.    The potential simplifies as:
\begin{equation}
V(v_{u}, v_{d}) = m_1^2 |v_{d}|^2 + m_2^2 |v_{u}|^2 - \mu B (v_u v_{d}) + c.c. + \frac{\bar{g}^{2}}{8} ( |v_{d}|^2 - |v_{u}|^2)^2.
\end{equation}
Note that we can use a ``Peccei-Quinn" rotation, under  the $H_{u}$ and $H_{d}$ fields rotate by the same phase to make $\mu B$ real and positive.
From here we can explicitly check that the vevs are in fact real, i.e. spontaneous breaking of $CP$ does not occur.  We define
\begin{eqnarray}
v_{u} = |v_{u}| e^{i(\alpha+ \beta)},\\
v_{d} = |v_{d}| e^{i(\alpha - \beta)}.
\end{eqnarray}
 Examining the potential as a function of the phases, we see that there is only a single term (the one proportional to $\mu B$) that depends on the overall phases of the vevs $\alpha$.  This contribution to the potential is minimized for $\alpha =0$.  The remaining relative phase $\beta$ can be removed by a $U(1)_Y$ gauge transformation -- recall the two doublets have opposite hyper charge.

From this potential, we should be able to derive the mass spectrum of the five physical Higgs boson states (there are eight degrees of freedom in two complex scalar doublets, and three are eaten by the $W^{\pm}$ and $Z$).  These states are denoted $h^{0}, H^{0}, H^{\pm}$ and $A^{0}$.  
 
But before doing this, note that there is a special direction where the quartic coupling vanishes.  The existence of these so-called ``flat-directions"  is a generic feature of supersymmetry gauge theories.  Since the coupling in question is generated by the $D$-terms, this direction is known as $D$-flat.  Its existence means we need to take care that the potential will not be unbounded from below.  To determine the condition  needed to prevent this instability, we set
\begin{equation}
v_{u} = v_{d} = \phi_{flat}.
\end{equation}
The potential in this direction is
\begin{equation}
V(\phi_{flat}) = \phi_{flat}^{2} (m_{1}^2 +m_{2}^{2} - 2 \mu B).
\end{equation}
This implies
\begin{equation}
\mu B < \frac{m_{1}^{2} + m_{2}^{2}}{2}.
\label{eqn:NoRunaway}
\end{equation}
On the other hand, to get EWSB (at all!), we need a negative mass eigenvalue at the origin ($v_{u}= v_{d} =0$).  One way to ensure this is the case is to require a negative determinant for the relevant mass matrix (recall 
$\lambda_{1} \lambda_{2} = \rm{det} {\mathcal  M}$, where $\lambda_{i}$ are the eignenvalues).  
\begin{eqnarray}
\rm{det} \left(\begin{array}{cc} 
\frac{\partial^{2} V}{\partial v_{u}^2} & \frac{\partial^{2} V}{\partial v_{u} \partial v_{d}} \\
\frac{\partial^{2} V}{\partial v_{u} \partial v_{d}} & \frac{\partial^{2} V}{ \partial v_{d}^{2}} 
\end{array} 
\right) < 0 &\Rightarrow&
\rm{det} \left(\begin{array}{cc} 
m_{2}^{2} & -\mu B\\
-\mu B &m_{1}^{2}
\end{array} 
\right) \\
& \Rightarrow& \mu B > m_{1}^{2} m_{2}^{2}.
\label{eqn:NegativeDet}
\end{eqnarray}
Note that if $m_{1}^{2}= m_{2}^{2}$, you cannot simultaneously satisfy Eqs.~(\ref{eqn:NoRunaway}) and (\ref{eqn:NegativeDet}).  This should not be viewed as a problem.  In fact, renormalization group evolution rapidly  suppresses $m_{H_{u}}^{2}$ (and drives it negative) due to its large coupling to the top quark. The ability to start with universal soft supersymmetry breaking masses, and see $m_{H_{u}}^{2}$ driven negative is viewed as a success of EWSB within the supersymmetric framework.  

Now we want to move on to discuss the spectrum of the Higgs bosons.  In much of the MSSM parameter space, there is a Higgs boson that is approximately Standard Model like.  We will try to work in a way that makes both its identity and its properties as explicit as possible. So, rather than making a full analysis of the Higgs potential in the $H_{u},  H_{d}$ basis (as is often done), we work in a basis that makes it manifest that one CP-even scalar, traditionally $h^{0}$, likely has properties that closely mimic those of the Standard Model Higgs boson.   We closely follow the presentation of Ref.~\cite{WellsHiggsNotes}, see also Ref.~\cite{Yeghian:1999kr,Dobrescu:2000yn} in another context.  This basis (dubbed the Runge basis in Ref.~\cite{WellsHiggsNotes} for no good reason), involves a rotation away from the $H_{u}, H_{d}$ basis to a basis where one field contains the full vacuum expectation value (vev).  In a sense it is this field that is \emph{the} Higgs boson.  Denote vacuum expectation values $\langle H_{d} \rangle \equiv v_{d}$,  $\langle H_{u} \rangle \equiv v_{u}$ and $v_{u} /v_{d} \equiv \tan \beta$, and 
\begin{equation}
\Phi_{vev} = \frac{v_{d}}{v} H_{d}^{c} + \frac{v_{u}}{v} H_{u},\\
\Phi_{\perp} = -\frac{v_{u}}{v} H_{d}^{c} + \frac{v_{d}}{v} H_{u}.
\end{equation}
Note that we have taken the conjugate of $H_{d}$, $H_{d}^{c} \equiv i \sigma^{2} H_{d}^{\ast}$, to make sure that we are adding fields with identical hypercharge.  \\


So, we can write:
\begin{equation}
\Phi_{vev} = \left(
\begin{array}{c}
G^{\pm} \\
\frac{1}{\sqrt{2}} ( v + h_{v}^0 + i G^{0})
\end{array}
\right)
\quad
\Phi_{\perp} =
\left(
\begin{array}{c}
H^{\pm} \\
\frac{1}{\sqrt{2}} ( H_{v}^0 + i A^{0})
\end{array}
\right),
\end{equation}
with $G^{\pm}, G^0$ the Goldstone bosons who get eaten to make $W^{\pm},Z^{0}$ massive.  Here we also clearly see  five extra degrees of freedom ($H^{\pm}, h^{0}_v, H^0_v, A^{0}$).  Although $H^{\pm},A^{0}$ are mass eigenstates, $h^{0}_v$ and $H^0_v$ are not.  $h^0_{v}$  is by construction aligned with the vacuum expectation value.  It is this field that is responsible for giving the gauge bosons their masses.  In the Standard Model, there is a single Higgs field (that must perforce be aligned with the vev), so we expect that $h^0_{v}$ will have identical couplings to the gauge bosons as a Standard Model Higgs.  
We expect 
any misalignment between $h^0_v$ and the mass eigenstate $h^{0}$ will result in non-Standard Model like couplings.

We can rewrite the original Higgs potential as
\begin{eqnarray}
V&=& (\mu^{2} + \sin^2{\beta} \, m_{H_{u}}^2 +  \cos^2{\beta} \, m_{H_{d}}^2 - \sin{2 \beta} B \mu) |\Phi_{vev}|^2 \nonumber \\
&+&  (\mu^{2} + \cos^2{\beta} \, m_{H_{d}}^2 +  \sin^2{\beta} \, m_{H_{u}}^2 + \sin{2 \beta} B \mu) |\Phi_{\perp}|^2 \nonumber \\
&+& \left( \frac{\sin{2 \beta}}{2} ( m_{H_{u}}^2 - m_{H_{d}}^2) - \cos{2 \beta} B \mu \right)  \left( \Phi_{vev}^{\dagger}  \Phi_{\perp} + c.c \right) \nonumber \\
&+& \frac{\bar{g}^{2}}{8} \left[ \cos 2 \beta \left(|\Phi_{\perp}|^2 - |\Phi_{vev}|^2 \right) + \sin {2 \beta} ( \Phi_{vev}^{\dagger} \Phi_{\perp} + c.c.) \right] \nonumber \\
&+& \frac{g^{2}}{2}|\Phi_{vev} \cdot \Phi_{\perp}|^2.
\end{eqnarray}
As usual, we can examine the minimum by taking derivatives and forcing them to be equal to zero.   Letting 
\begin{equation}
\phi_{i} = \left\{ G^{\pm}, H^{\pm}, G^{0}, h_{v}^{0}, H_{v}^{0}, A^0 \right\},
\end{equation}
we have:
\begin{equation}
\left.\frac{\partial V}{\partial \phi_{i}} \mbeq 0 \right|_{\{h=v, \phi_{i} \neq h=0\}  \equiv minimum}
\end{equation}
It is a straightforward exercise to check that only
\begin{equation}
\left.\frac{\partial V}{\partial h_{v}}\mbeq 0 \right|_{minimum}
\quad
\left.\frac{\partial V}{\partial H_{v}} \mbeq 0 \right|_{minimum}
\end{equation}
yield non-trivial conditions:
\begin{eqnarray}
h_v&:& \;  \frac{\bar{g}^2}{8} v^{2} \cos^2{2 \beta}+  \mu^{2} + m_{H_{d}}^{2} \cos^2{\beta} + m_{H_{u}}^{2} \sin^2{\beta} - B \mu \sin{2 \beta}  =0, \label{eqn:hvcond}\\
H_v&:& \;  -\frac{\bar{g}^2}{8} v^{2} \cos{2 \beta} \sin {2 \beta}+ \frac{1}{2}( m_{H_{u}}^{2}- m_{H_{d}}^2) \sin{2 \beta}  - B \mu \cos{2 \beta}  =0. \label{eqn:Hvcond}
\end{eqnarray}

Taking linear combinations (and using $M_{Z}^2= \bar{g}^{2} v^{2}/4)$, we find:
\begin{eqnarray}
\label{eq:EWSBtune}
M_{Z}^{2} + \mu^{2} = \frac{m_{H_{d}}^{2} - \tan^{2}{\beta} m_{H_{u}}^2}{\tan^2{\beta}-1},\\
\frac{B \mu}{\sin{2 \beta}} = \mu^{2} + \frac{1}{2} \left( m_{H_{d}}^{2} + m_{H_{u}}^{2} \right).
\end{eqnarray}
These two conditions should be familiar to MSSM aficionados, even though the route we have taken may not be.

Before moving on to discuss the phenomenology of the Higgs bosons, it is worthwhile to pause and consider the implications of this equation for fine-tuning.  The weak scale (i.e. $M_Z$) is determined in terms of $\mu$, $m_{H_{d}}^{2}$, and $m_{H_{u}}^{2}$.  We already pointed out that the absence of fine-tuning will require relatively light stops and gluinos.  Here, we see clearly why:   if they are heavy, it puts pressure on $m_{H_{u}}^2$ and in turn $M_{Z}$.  There is a new lesson, too.  The $\mu$ term cannot be too large.  If it is, there is already fine tuning at the \emph{tree} level in Eq.~(\ref{eq:EWSBtune}).  So, it seems that a natural spectrum requires that Higgsinos not be too heavy, either.

From here, to get the Higgs boson masses, we just need to take the potential, and 
\begin{enumerate}
\item{Expand in terms of $\left\{ v, h_{v}^0, H^{\pm}, H_{v}^{0}, A^{0} \right\}$.}
\item{Use conditions (\ref{eqn:hvcond}) and (\ref{eqn:Hvcond}) above.}
\item{Pick off terms quadratic in the fields, i.e. the mass terms.}
\end{enumerate}

The results are particularly simple for $A^{0}$ and $H^{\pm}$ because they do not mix with other fields.
The pseudoscalar Higgs boson's mass is given by  
\begin{equation}
M_{A}^{2} = ({m}_{H_{u}}^{2} -{m}_{H_{d}}^{2}) \sec{2 \beta} - M_{Z}^2.
\end{equation}
The interested reader can check that this expression is proportional to $B \mu$ -- the $A$ is a Goldstone boson of the PQ breaking in the limit that $B\mu$ vanishes.  The charged Higgs mass is given  by
\begin{equation}
M_{H^{\pm}}^2=M_{A}^{2} +M_W^2.
\end{equation}

For CP-even neutral Higgs bosons,we have to diagonalize the $2\times2$ $h^{0}_v$ and $H^{0}_v$ mass matrix given by
\begin{equation}
{\mathcal M}^2_{hH}=
\left(\begin{array}{ll}
M_Z \cos^{2}{2 \beta} & \; \; \; \; -M_{Z}^2 \sin{2 \beta} \cos{2 \beta} \\
-M_{Z}^2 \sin{2 \beta} \cos{2 \beta} & \; \; \; \; (\tilde{m}_{H_{u}}^{2} -\tilde{m}_{H_{d}}^{2})  - M_Z^{2} \cos^{2}{2 \beta}
\end{array}
\right)
\end{equation}
The $(1,1)$ entry of the mass matrix (corresponding to the $h^{0}_{v}$ state) has $(mass)^2 = M_{Z}^2 \cos^{2}{2 \beta}$.  A theorem from linear algebra states that the smallest eigenvalue of a matrix is smaller than the smallest diagonal element of that matrix. Trigonometry then indicates a Higgs mass less than $M_{Z}$.   Furthermore, this is saturated as $(\tilde{m}_{H_{u}}^{2} -\tilde{m}_{H_{d}}^{2}) \rightarrow \infty$, where $h^{0} \equiv h$.   But the Higgs boson is (apparently) at 125 GeV.  How robust is this prediction?  After all, we have not yet taken supersymmetry breaking into account.  We do not expect loop effects to cancel once supersymmetry is broken \cite{Haber:1990aw,Ellis:1990nz,Okada:1990vk}.

There are many ways to compute these effects.  They include 
\begin{enumerate}
\item{A direct diagrammatic calculation}
\item{A renormalization group analysis}
\item{Computation of the effective potential.  (For a textbook treatment, see Sec. 10.6 of Ref.~\cite{DreesText}).}
\end{enumerate} 

We concentrate on the second of these --  in some ways it offers the most physical insight.  We take the scale of supersymmetry breaking $\tilde{m} >> M_{Z}$, we can imagine matching the full Minimal Supersymmetric Standard Model onto the Standard Model at that scale.  Above $\tilde{m}$  supersymmetry enforces the equality of the quartics and the gauge couplings.  Because the Higgs quartic is tied to gauge couplings in the supersymmetric regime, at scales above the supersymmetry breaking scale $\tilde{m}$, we have
\begin{eqnarray}
m_h^2 &=&  2 \lambda v^{2} \nonumber \\
&=&2\left(\frac{1}{8} (g^{2} + g_{Y}^2)\right) v^{2} .
\label{eqn:HiggsRelation}
\end{eqnarray}
Below $\tilde{m}$ the quartic is no longer tied to the gauge coupling -- the symmetry no longer connects them, and they are free to evolve independently.  For scales not \emph{too} far from $\tilde{m}$ we have (with $t \equiv \log{Q}$)

\begin{figure}[t]
\includegraphics[width=.8\textwidth]{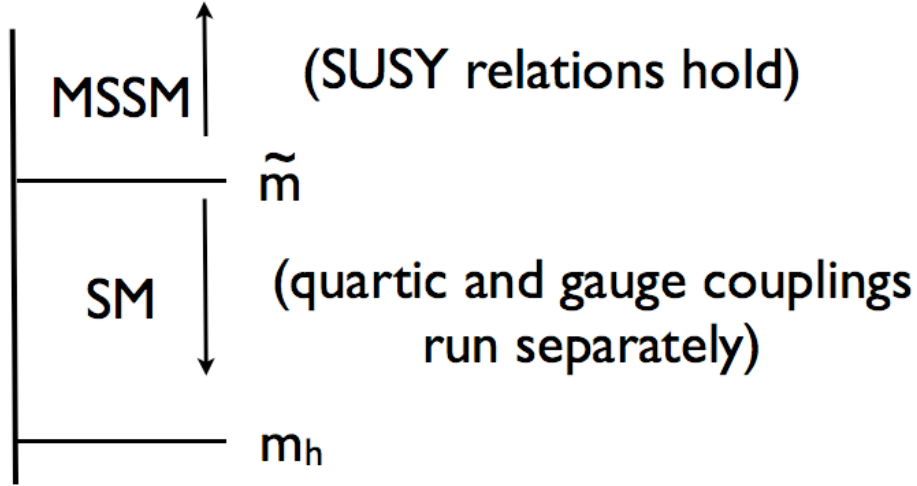}
\caption{Matching in a  calculation of the mass of the Higgs boson}
\label{fig:HiggsMatch}
\end{figure}

\begin{eqnarray}
\lambda(Q) &=& \lambda(\tilde{m})  + \frac{ \partial \lambda}{\partial t} (t-t_{0}) + \ldots \\
&=& \lambda(\tilde{m}) - \beta_{\lambda} \log{\left(\frac{\tilde{m}}{Q}\right)}.
\end{eqnarray}
Here $\beta_{\lambda}$ is the beta-function for the quartic in the Standard Model.  The leading contribution (arising from the top quark Yukawa coupling) is 
\begin{equation}
\beta_{\lambda} = -\frac{3 y_{t}^{4}}{8 \pi^2}.
\end{equation}
Setting $Q= m_t$, and using Eq.~(\ref{eqn:HiggsRelation}), we find
\begin{equation}
m_h^2 = M_{Z}^2 \cos^{2} 2 \beta +\frac{3 g^{2} m_{t}^4}{4 \pi^2 M_{W}^2} \log{\left( \frac{\tilde{m}}{m_{t}}\right)}
\end{equation}
For a Higgs boson mass of 125 GeV, this indicates a supersymmetry breaking scale of approximately 1 TeV.  

This is the core of the argument that supersymmetry is now at least somewhat fine tuned.  The above line of reasoning indicates pressure for the top squarks to be heavy.  On the other hand, the heavier the stops get, the more fine-tuned the Higgs mass squared parameter is.    So, even if limits from the LHC had not been strengthening the bounds on superpartners, the Higgs boson mass would have been a clue that the superpartners are likely somewhat heavy.

It should noted that the above formula is a simplification.  In particular, it neglects the possible contribution of the trilinear scalar term  $A_{t} \tilde{q_{3}} \tilde{u^{c}_3} h$ to the Higgs mass.  Indeed, pushing $A_{t}$ to be somewhat large (its effects are maximized at $A_{t} \sim \sqrt{6} \tilde{m_{t}})$ can have an important effect on the Higgs boson mass.  This does not come for free:  while (finite) diagrams with the $A$-term can push the effective value of the quartic (and hence the Higgs mass) higher, the same $A$ terms will contribute a radiative correction to the Higgs mass squared parameter, and thus the fine-tuning, see Eq.~(\ref{eqn:stoploopcorr}).

In fact, if the supersymmetry breaking scale is not ``too high'', then one is forced to consider relatively large $A$-terms .  What implications does this have for the underlying theory?  It is interesting to note that in its simplest forms of so-called ``gauge mediated supersymmetry breaking" do not communicate large $A$-terms.  However, $A$ terms can be generated via renormalization group evolution from a high mediation scale (where the messenger particles lie) down to the TeV scale.  As recently reviewed in Ref.~\cite{Draper:2011aa}, this indicates that gauge mediated versions of the MSSM will necessarily possess relatively heavy squarks, high mediation scales, or both. 

\section{What does 125 GeV mean?}
\subsection{How uncomfortable is 125 GeV?}
So, is 125 GeV ``natural"?  It is borderline.   The Higgs boson could have been found at 160 GeV.  To achieve such a large value would have required extraordinarily fine-tuning in an supersymmetric theory.  In fact,  a theory that is the MSSM above $m_{GUT} \simeq 10^{16}$ GeV and the Standard Model below predicts a Higgs boson closer to 140 GeV \cite{Hall:2009nd}.  But nature didn't choose that path.  On the other hand, nature could have chosen to place $m_{h}$ right near $M_{Z}$, with minimal contribution to the quartic from supersymmetry breaking.  In this case radiative corrections would have been minimal.  But that ship sailed some time ago.   125 GeV is tantalizingly close to $M_{Z}$, but just far enough away to give us  pause.
Where does that leave us?
\begin{itemize}
\item{ One approach  (see below)  is to ask for new physics to affect the Higgs boson mass.  If the low energy theory is not just the Higgs sector of the MSSM, it is possible that there are new contributions to mass of the the Standard Model-like Higgs boson.}
\item{The alternative is to accept some (or a lot, see Sec.~\ref{sec:split}) of tuning.   Supersymmetry was designed to get rid of a tuning of some thirty-two orders of magnitude.  How uncomfortable would we be with a 1\% accident in the fundamental theory of our universe?  How about a part in $10^{4}$?  At what point do we stop thinking of supersymmetry as a solution to a hierarchy problem?}
\item{Supersymmetry is ``just around the corner", but the Higgs mass is achieved through particularly large $A$-terms.  There a designer spectrum that gives rise to a large Higgs, and we have eluded the LHC for now, but won't for much longer.}
\end{itemize}


\subsection{Higgs Properties Implications}
At present, there is no strong deviation of the observed ``Higgs boson'' properties from that of the Standard Model Higgs boson.  This is already quite an interesting statement for supersymmetry.  In the MSSM, it would have been possible (via $h^{0}_v$-$H_v^0$ mixing) to modify the properties of the Higgs boson significantly.  Though it is early days, the apparent absence of deviations indicates that we may lie in the so-called ``decoupling regime'', where $M_A \rightarrow \infty$ and the other Higgs bosons decouple from the theory.

The only deviation worth mentioning at present -- and it is a mild one whose significance is unclear --  is a slight enhancement in the rate  $R_{\gamma \gamma}$: the number of Higgs bosons decaying to two photons is higher than expected from the Standard Model by  roughly $1.8 \pm .4$ (ATLAS \cite{ATLASgamma}) and $1.6 \pm 0.5$ (CMS \cite{CMSgamma}).  This is tantalizing because the decay to two photons is induced in the Standard Model via loops of W-bosons and to a lesser extent top quarks.  This raises the possibility that there could be new particles with weak scale masses (very roughly at the few 100 GeV scale or below if they are to have a large effect) that could also be running around the loop, and modifying the branching ratio.  This has been a subject of much theoretical speculation in a variety of theoretical frameworks.  The best suspect that could conceivably be responsible for modifying this branching ratio within the MSSM is the stau (the superpartner of the  $\tau$ lepton).  Because it has large $|\rm{charge}|=1$ it has a large coupling to photons, and because it lacks color, it would not disturb the $gg \rightarrow h$ rate, which might have shown up, e.g. in the overall rate for the process $gg \rightarrow ZZ^{\ast}$ (this disfavors the stops as the culprit).  Depending on the stau trilinear coupling, the staus can in fact cause the branching ratio to go in either direction.   A recent analysis of this can be found in Ref.~\cite{Carena:2011aa,Carena:2012gp};  it is too early to draw conclusions from the experimental data.

\subsection{Beyond the MSSM?}
The MSSM has a curious feature.  In its superpotential, there is a term, 
\begin{equation}
W = \mu H_{u} H_{d}
\end{equation}
with positive mass dimension.  What should the size of this $\mu$ term be.  From our considerations of electroweak symmetry breaking in the previous section, we see that, phenomenologically, it must be of order $M_{Z}$ to realize electroweak symmetry breaking without fine-tuning.  This presents a puzzle.  After all, terms in the superpotential are, by construction {\it supersymmetric}.  They have nothing to do with the scale of supersymmetry breaking.  

A natural extension to the MSSM is to introduce a single electroweak singlet superfield, $S$. The part of the superpotential that gives rise to the Yukawa couplings remains unchanged, but the piece that previously gave rise to the $\mu$ term is replaced:
\begin{equation}
W_{\mu} = \mu H_{u} H_{d} \rightarrow W_{NMSSM}=  \lambda S H_{u} H_{d} + \kappa S^{3}
\end{equation}
If $S$ gets a vacuum expectation value, an effective $\mu$ term is generated with $\mu = \lambda \langle S \rangle$.  It might appear that no progress has been made -- all that has been done is to replace an unexplained value of $\mu$ with an unexplained value of $\langle S \rangle$.  However, it is reasonably straightforward to arrange for $\langle S \rangle$ to vanish in the supersymmetric limit, in which case it is plausible that $\mu_{eff}$ is of order the weak scale.  One possibility is to have a negative (mass)$^2$ term for $S$ coming from supersymmetry breaking that drives the vev.

Interestingly, this superpotential also gives a novel contribution to the mass of the Higgs boson.  In particular, the first term contributes at tree level
\begin{equation}
\left. \Delta m_{h}^{2}\right|_{NMSSM} = \frac{\lambda v^{2}}{2} \sin {2 \beta}
\end{equation}
with $v = 246$ GeV.  This term gives the possibility of raising the Higgs boson mass without a large supersymmetry breaking contribution.  It should be noted that this term is maximized precisely only where the tree-level MSSM  ($D$-term) contribution is minimized.  So, this is no panacea when it comes to getting a heavy Higgs boson.  Furthermore,  the maximum value of $\lambda$ is limited if one imposes that there is no Landau pole (and no new physics) below the grand unified scale: $\lambda < 0.7$.    Because one extraordinarily nice feature of the MSSM was its unification, hitting a Landau pole before the couplings had a chance to unify would be a pity.   As recently reviewed \cite{Hall:2011aa}, large values of $\lambda$ and small values of $\tan \beta$ can allow relatively light stops, with masses of roughly 500 GeV.

\section{Conclusion}
No direct hints for supersymmetry have been found. The recent discovery of a Higgs boson, however, at 125 GeV gives us some important tea leaves to read.  Because supersymmetry relates the Higgs boson mass to the $Z$-boson mass, the discovery of a ``light'' Higgs boson, i.e. near 100 GeV, is perhaps encouraging.  On the other hand, 125 GeV is not 91 GeV.  What should we make of the discrepancy?

Perhaps supersymmetry is broken more strongly than would would have expected based on naive fine-tuning arguments.  One interesting possibility is one whether the fermionic superpartners are found near the TeV scale, but the superpartners are parametrically a loop heavier, near the 100 TeV scale.  This can realize the Higgs boson mass rather simply, but raises sharp questions about naturalness.  Supersymmetry will have softened the hierarchy problem, but the theory at the weak scale will be tuned to a part in a $10^5$ or so.  From a bottom-up perspective, this is puzzling, but perhaps there are motivations from a more fundamental theory. 

Another possibility is that the theory at the weak scale is not quite the MSSM, but rather has some twist, e.g. the NMSSM or beyond.   These extensions can help raise the Higgs boson mass, but one still must address the absence of direct production of superpartners.  

Of course it is possible that supersymmetry is not important for physics at the weak scale at all.

The upgrade of the LHC will have a great deal to say about whether there are in fact any superpartners near the TeV scale.  Information from experiments searching for weakly interacting massive particles (WIMPs) via either direct or indirect detection are also rapidly approaching the sensitivities relevant for the MSSM.  It is an exciting time, and if I were to give similar lectures a few years from now, my bet is that they would look very different.

\section{Acknowledgments and Appendices}
This work was supported in part by the DOE Office of Science under Grant DE-SC0007859 and NSF CAREER Grant NSF-PHY-0743315.  Thanks to the Berkeley Center for Theoretical Physics for their hospitality while these notes were completed. Thanks to the students of TASI 2012 for good questions and a stimulating atmosphere.  These notes also owe a debt to Nima Arkani-Hamed, Savas Dimopoulos, Lawrence Hall, Gordy Kane, Hitoshi Murayama, Jay Wacker, and James Wells,  who have impacted my thinking on these topics through courses, collaborations, and conversations.  Thanks also to Jack Kearney for a reading of this manuscript. All errors, are of course, my fault.


\bibliographystyle{ws-procs9x6}
\bibliography{PierceTASI}

\end{document}